\documentclass[aps,twocolumn,showpacs,amsmath,superscriptaddress,amssymb,longbibliography,prb,10pt]{revtex4-2}
\usepackage[colorlinks=true,allcolors=blue,hypertexnames=false]{hyperref}
\usepackage{amsmath}
\usepackage{dsfont}
\usepackage{hyperref}
\usepackage{textcomp}
\usepackage{tabularx}
\usepackage{graphicx}
\usepackage{soul}
\usepackage{feynmp-auto}
\usepackage{footmisc}
\usepackage[export]{adjustbox}
\usepackage{wrapfig}
\usepackage{bbm}

\makeatletter
\def\@bibdataout@aps{%
 \immediate\write\@bibdataout{%
  @CONTROL{%
   apsrev42Control%
   \longbibliography@sw{%
    ,author="48",editor="1",pages="0",title="0",year="1"%
   }{%
    ,author="48",editor="1",pages="0",title="",year="1"%
   }%
  }%
 }%
 \if@filesw
  \immediate\write\@auxout{\string\citation{apsrev42Control}}%
 \fi
}%
\makeatother

\newcommand{\beg}{\begin{equation}}
\newcommand{\en}{\end{equation}}
\newcommand{\bp}{\mathbf p}
\newcommand{\bq}{\mathbf q}

\newcommand{\bk}{\mathbf k}
\newcommand{\br}{\mathbf r}

\newcommand{\bn}{\mathbf n}

\newcommand \bel  {\begin{align}}
\newcommand \enl  {\end{align}}

\newcommand{\eps}{\epsilon}
\newcommand{\veps}{\varepsilon}

\def\XXint#1#2#3{{\setbox0=\hbox{$#1{#2#3}{\int}$}
     \vcenter{\hbox{$#2#3$}}\kern-.5\wd0}}

\begin{document}

\title{Finite-momentum coupling of Higgs and Bardasis--Schrieffer modes in
superconductors with competing pairing channels}

\author{Samuel Awelewa}
\affiliation{Department of Physics, Kent State University, Kent, Ohio 44242, USA}

\author{Yafis Barlas}
\affiliation{Department of Physics, University of Nevada, Reno, Nevada 89557, USA}

\author{Maxim Dzero}
\affiliation{Department of Physics, Kent State University, Kent, Ohio 44242, USA}

\date{\today}

\begin{abstract}
In superconductors with competing pairing channels, two well defined excitations
exist below the pair-breaking edge: the Higgs mode of the condensed
$s$-wave channel and the Bardasis--Schrieffer (BS) exciton of the
subdominant $d$-wave channel. Their mixing is doubly forbidden --- by
point-group symmetry at zero momentum and because the two reside in
the amplitude and phase sectors of the order parameter respectively, by
particle--hole symmetry at every momentum. Working in a Nambu--Keldysh
quasiclassical framework extended to leading $1/\varepsilon_F$
corrections and including the self-consistently screened Coulomb
potential, we show that finite momentum combined with particle--hole
asymmetry generates a direct coupling which we obtain in closed form. Whether this coupling produces an avoided crossing
is decided, however, not by its magnitude but by kinematics. In the
clean limit the Higgs is not a sub-gap pole but a resonance pinned to
the pair-breaking edge, which disperses with coefficient unity in
$(v_Fq)^2$, while the bound BS mode disperses more slowly: the two
branches therefore separate rather than converge and never become
degenerate. The coupling instead redistributes spectral weight: the BS pole acquires
a small amplitude (Higgs) component, producing a line in the amplitude
channel at the BS frequency, where that channel otherwise has strictly
zero sub-gap response. This induced weight grows as $q^2$ with a
$\cos2\phi_{\mathbf q}$ angular dependence, vanishing at zero momentum
and along the nodal direction, while the BS mode itself remains a
sharp sub-gap pole at every momentum. The obstruction is specific to the clean limit: exact
dirty-limit results show that disorder detaches the amplitude resonance
from the edge and reverses its dispersion, which can result in an avoided crossing with the BS mode at intermediate scattering. 
In that regime, the
coupling computed here would set the splitting between the hybridized
branches. We
discuss the experimental implications of these results.
\end{abstract}

\maketitle
\section{Introduction}
Collective modes of the superconducting condensates encode the structure of the pairing
interaction beyond what the equilibrium gap reveals. The amplitude (Higgs) mode of the
condensate~\cite{VolkovKogan1974,LittlewoodVarma1981,LittlewoodVarma1982,PekkerVarma2015, ShimanoTsuji2020,Shimano2013,Matsunaga2017, Katsumi2018,SooryakumarKlein1980, Measson2014} and the phase mode, promoted to the
plasma frequency by the long-range Coulomb interaction~\cite{Anderson1958,Anderson2015,CarlsonGoldman73,Volkov1975,Volkov1979,SchmidSchon1975,SchmidSchon1979,Kulik1981}, are
properties of the condensed channel itself. A third family of modes exists whenever the
pairing interaction contains a subleading channel of different symmetry. For example, in their seminal work Bardasis and
Schrieffer~\cite{BardasisSchrieffer1961} showed that in an $s$-wave superconductor with
a subdominant attraction of higher angular momentum, pair fluctuations into the
unoccupied channel form an exciton-like bound state below the pair-breaking edge
$2\Delta$. The energy of this Bardasis--Schrieffer (BS) mode measures the proximity of
the subleading channel to its own condensation point, softening to zero at the boundary
of the time-reversal symmetry breaking $s+id$ state: the mode is, in effect, a
spectroscopic ruler for the competition between pairing channels. For example, electronic Raman
scattering has resolved BS modes in
Ba$_{0.6}$K$_{0.4}$Fe$_2$As$_2$~\cite{Kretzschmar2013,Bohm2014}, establishing a strong
subdominant $d_{x^2-y^2}$ channel in the iron pnictides and turning the mode from a
theoretical curiosity into a quantitative probe of the pairing
hierarchy~\cite{Maiti2016}. Very recently, terahertz nonlinear spectroscopy has reported
a subgap collective resonance of BS character in FeSe~\cite{FeSeTHz2025}, extending the
experimental tools of probing collective modes beyond Raman spectroscopy.

On the theory side, the collective-mode landscape of superconductors with competing
$s$- and $d$-wave interactions has been mapped in considerable detail at zero momentum. Specifically,
Maiti and Hirschfeld~\cite{MaitiHirschfeld2015} classified the phase, amplitude, BS, and
mixed-symmetry BS modes across the full $s\to s+id\to d$ phase diagram for one- and
two-band models. Nonequilibrium studies established how these modes ring after external electromagnetic field has been applied. Recently,
M\"uller, Volkov, Paul, and Eremin~\cite{Muller2019} showed that pulse polarization and
intensity select whether the Higgs or the BS mode is excited, with pronounced BS
signatures in the quasiparticle distribution. The interplay of nematicity with BS
modes in short-time dynamics was analyzed in Ref.~\cite{MullerEremin2021}. Inside the
time-reversal symmetry breaking condensed phases, Neri, Metzner, and
Manske~\cite{Neri2025} classified the amplitude, relative-amplitude, and relative-phase
oscillations of $\Delta_1+i\Delta_2$ condensates and matched them to symmetries of external field.
One should note that in all of these works the collective modes live at $\bq=0$: the photon momentum enters,
if at all, only as a selection device.

A parallel line of work asks how these modes can be made visible at all, since most of
them remain dark in far-field linear response. One route to bring these modes to light is to include the dispersion of the BS and amplitude modes \cite{ChubukovHiggs2023,Nosov2025}. Recently, Sun, Fogler, Basov,
and Millis~\cite{SunMillis2020} showed that at $\bq\neq0$, accessible to terahertz
near-field optics, the BS mode acquires a direct coupling $\propto\xi q$ ($\xi=v_F/\Delta$ is the superconducting coherence length) to the
longitudinal response --- generically stronger than that of the Higgs mode, which
couples only through particle-hole asymmetry --- and that both modes anticross with the
low-lying plasmon of a layered system (see also \cite{Sun2020}). Considering a superconductor which carries a supercurrent offers another route. For example,
Niederhoff, Kataoka, Takasan, and Tsuji~\cite{Niederhoff2025} derived the current-enabled linear
optical conductivity of single- and multiband BS, mixed-symmetry BS, and Leggett modes,
the condensate momentum playing the role that finite $\bq$ or the probe geometry plays
elsewhere. The momentum-resolved structure of the phase and amplitude sectors of a
$d$-wave superconductor, including the long-range Coulomb interaction, was recently worked out
in Ref.~\cite{Kazi2026}.

\begin{figure}[t]
\includegraphics[width=0.975\linewidth]{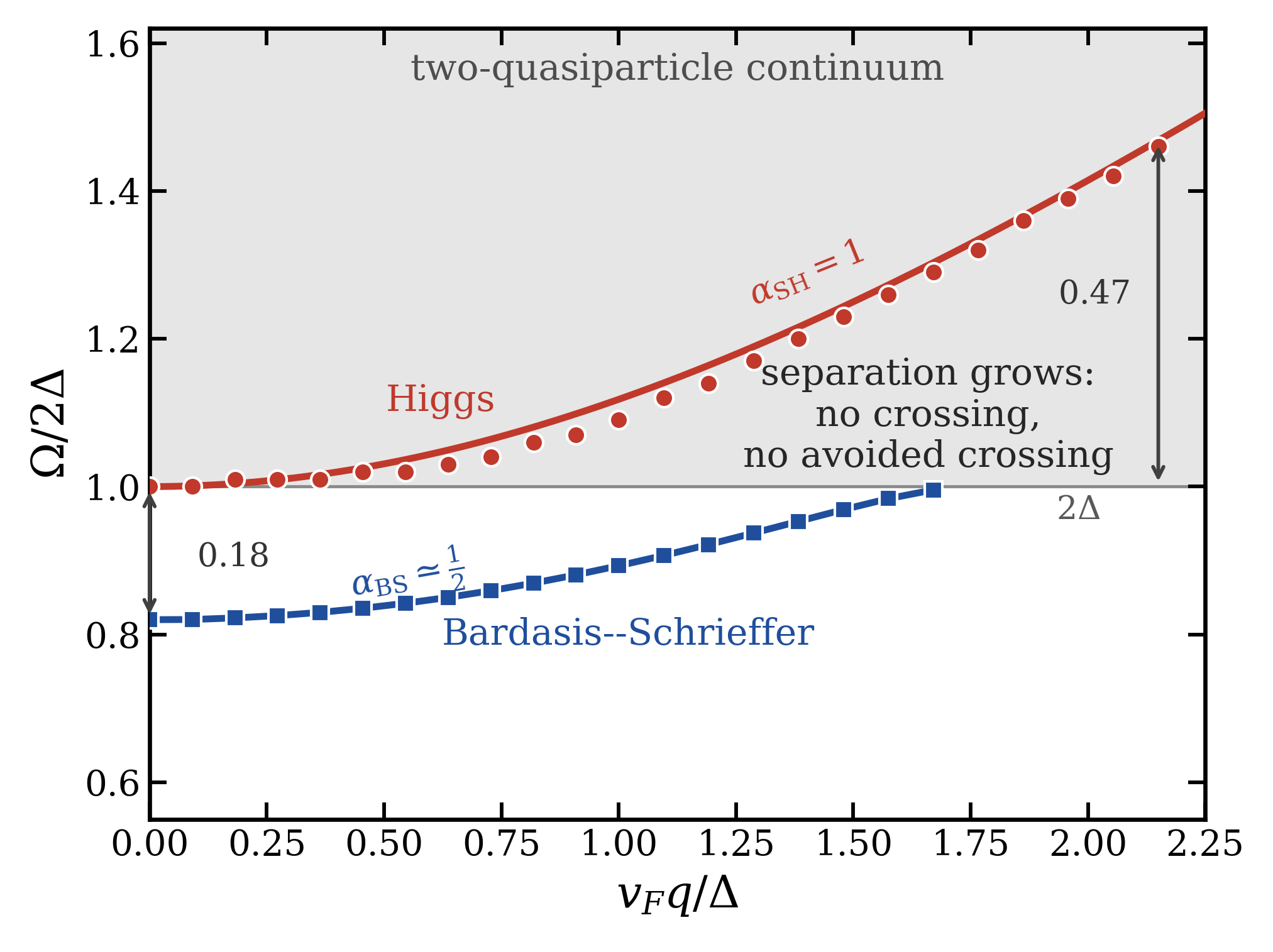}
\caption{\footnotesize Computed dispersion of the Higgs and
Bardasis--Schrieffer (BS) branches in a clean superconductor with a dominant
$s$-wave ($A_{1g}$) and a subdominant $d$-wave ($B_{1g}$) pairing channel,
for $\bq$ along the antinodal direction (Ba-122 calibration of $\lambda_d$, see Sec. IV,  was used). Red points are the maxima of the
amplitude spectral weight $A_{11}$ obtained from inverting matrix $\hat M$
defined by \eqref{LargeHiggsBS}; blue squares are the sharp sub-gap BS pole
obtained from $\det\hat M=0$ (below $2\Delta$ every entry of $\hat M$ is
real, so the pole position coincides with the $\delta$-function line of
$A_{44}$, and there is no linewidth to display). The grey region is the
two-quasiparticle continuum, whose bottom lies at $2\Delta$ for every
$\bq$. $A_1$ carries no spectral weight below it. The Higgs possesses no
sub-gap pole: its maximum follows the pair-breaking edge
$\Omega_{\mathrm{edge}}^2=(2\Delta)^2+(v_Fq)^2$ (red curve), which
disperses with unit coefficient, $\alpha_{SH}=1$, while the bound BS mode
disperses with $\alpha_{BS}\simeq1/2$ [notation of Eqs.~\eqref{nogo} and
\eqref{qstar}; contrast Fig.~\ref{fig:disorder}, where disorder softens
$\alpha_{SH}\simeq0$]. Because of this, the separation between the
branches \emph{grows} with momentum, from $0.18$ to $0.47$ in the units of
$2\Delta$ over the range shown; beyond $v_Fq\simeq1.7\Delta$ the BS
binding falls below our frequency resolution,
$\Omega_{BS}\to2\Delta^-$. The branches never become degenerate and there
is no avoided crossing.} 
\label{Fig1-Schematic}
\end{figure}
\begin{figure}[t]
\includegraphics[width=0.975\linewidth]{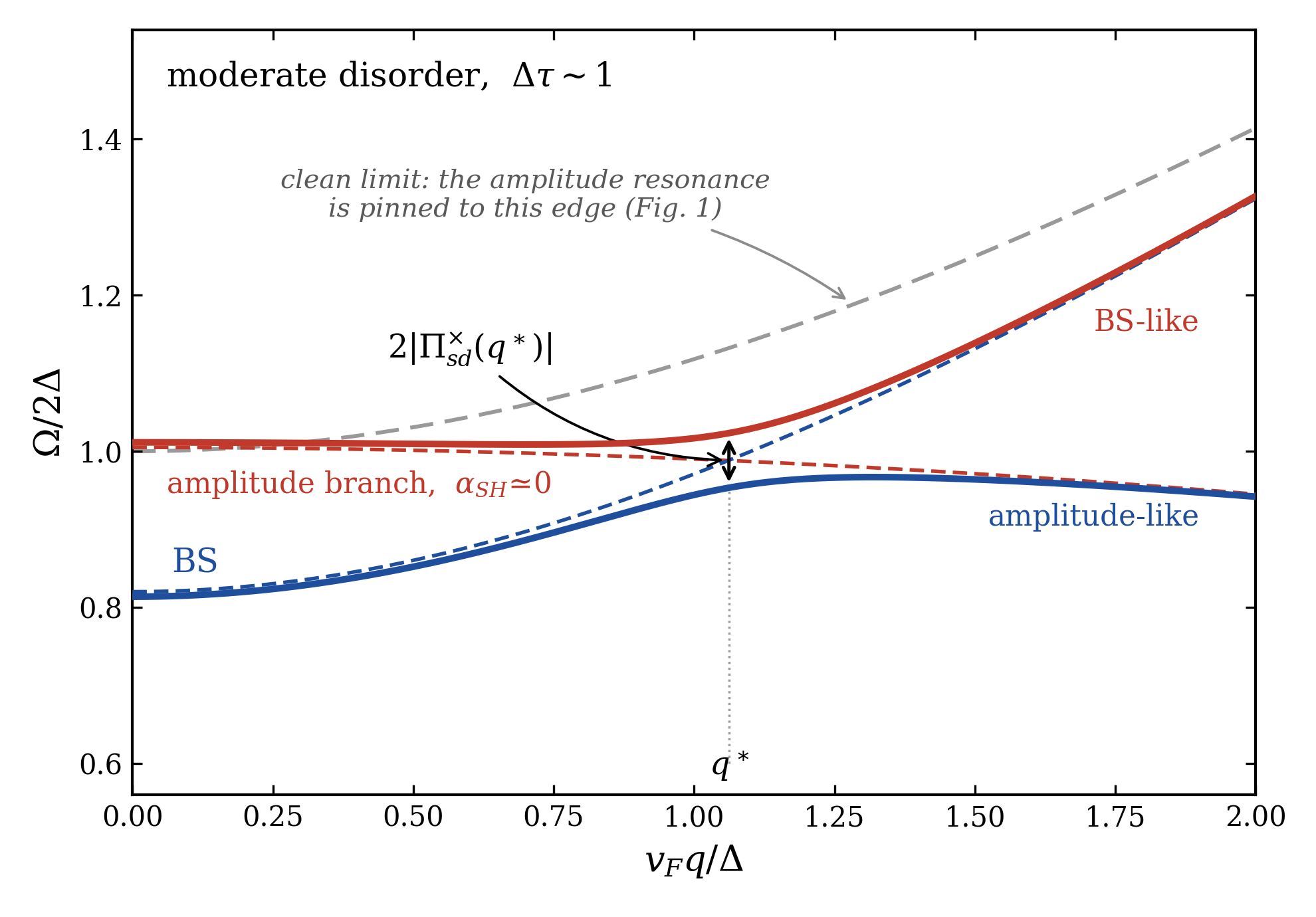}
\caption{\footnotesize {Schematic} of the anticipated moderate-disorder
scenario, $\Delta\tau\sim1$; this figure is illustrative and not a result of
the present calculation. Disorder detaches the amplitude resonance from the
pair-breaking edge and softens its dispersion through zero, so that
$\alpha_{SH}\simeq0$ (dashed red) while the BS continues to disperse upward
(dashed blue). The crossing condition \eqref{qstar} then has a solution at $q^\ast$, and the coupling computed in this work opens a gap
$2|\Pi^{\times}_{sd}(q^\ast)|$ between the hybridized branches (solid), which
exchange character across it. For contrast, the grey dashed curve is the
clean-limit pair-breaking edge of Eq.~\eqref{edge}: in that limit the
amplitude resonance is pinned to it, $\alpha_{SH}=1$ (i.e. $\alpha_{SH}>\alpha_{BS}$),
Eq.~\eqref{qstar} has no solution, and the branches separate rather than
cross (Fig.~\ref{Fig1-Schematic}).}
\label{fig:disorder}
\end{figure}
To the best of our knowledge, one major aspect of the physics of the collective modes which has not been addressed yet is 
the direct interaction of the two sharp
sub-$2\Delta$ excitations themselves: the Higgs mode of the dominant channel ($s$-wave in our case) and the BS
mode of the subdominant one ($d$-wave). At $\bq=0$ their mixing is forbidden by spatial symmetry ---
they belong to different irreducible representations - $A_{1g}$ versus $B_{1g}$ - of the
tetragonal point group. Less widely appreciated is the fact that finite momentum alone does
{not} lift the symmetry obstruction. The BS mode is the out-of-phase (transverse)
fluctuation of the subdominant order parameter, while the Higgs is an in-phase
(longitudinal or amplitude) fluctuation. Within a particle-hole symmetric theory --- and the standard
quasiclassical approximation is particle-hole symmetric by construction --- the
amplitude and phase sectors decouple at \emph{all} momenta. Finite $\bq$ does generate
intra-sector mixing with the characteristic $(v_Fq)^2\cos2\phi_{\mathbf q}$ angular
structure, but the finite-$\bq$ partner of the Higgs in the amplitude sector is the
overdamped in-phase $d$-wave fluctuation, which has no subgap pole, and the finite-$\bq$
partner of the BS mode in the phase sector is the plasmon. This is the anticrossing already
described in Ref.~\cite{SunMillis2020}. A genuine Higgs--BS hybridization therefore
requires two ingredients simultaneously: a finite probe momentum, to lower the point
symmetry, and particle-hole asymmetry, to connect the amplitude and phase sectors.

In this work we address this problem by directly evaluating the coupling
between the Higgs and BS modes. We find the following three results: (i) the
coupling itself; (ii) a kinematic obstruction that prevents it from producing
a resonance in a clean superconductor, Fig.~\ref{Fig1-Schematic}, and (iii) the
identification of the regime --- moderate disorder --- in which the
obstruction is lifted, and a genuine avoided crossing should appear,
Fig.~\ref{fig:disorder}.

To compute the coupling we use the Keldysh quasiclassical framework extended by the leading particle--hole asymmetric terms of order $O(\Delta/\varepsilon_F)$
(here $\Delta$ is the equilibrium pairing gap and $\varepsilon_F$ the
Fermi energy) to analyze 
the coupled four-component order-parameter
fluctuations --- amplitude and phase in both the $s$ and $d$ channels --- together with the fluctuation associated with the presence of the Coulomb potential. The
resulting Higgs--BS coupling is symmetry-protected and takes the form
$\Pi^{\times}_{sd}(\bq,\Omega)\propto(\Delta/\varepsilon_F)(v_Fq)^2\,
\Omega\,\cos2\phi_{\mathbf q}$: it switches on quadratically with
momentum, closes along the nodal directions of the subdominant gap, and
is odd in frequency, as an amplitude--phase coupling must be. We obtain
it in closed form, including its exact frequency profile, and confirm
each of these features independently.

Having the coupling in hand, one is led to ask whether it produces a
resonance --- an avoided crossing where the dispersing BS branch meets
the Higgs. In the clean limit it does not, and the reason is kinematic
rather than dynamical: it follows from the dispersions alone and is
independent of the size of the coupling. The Higgs of a clean $s$-wave
superconductor is not a pole of the longitudinal pair susceptibility
below the gap but a resonance pinned to the pair-breaking threshold. We
find that its spectral weight $A_1(q,\Omega)$ vanishes identically for
$\Omega<2\Delta$ and that the threshold disperses as
\beg\label{edge}
\Omega^2_{\rm edge}=(2\Delta)^2+(v_Fq)^2, 
\en
with coefficient exactly
unity. The BS mode, being a state bound below that continuum, disperses
as $\Omega^2_{BS}(q)=\Omega^2_{BS}(0)+\alpha\,(v_Fq)^2$ with $\alpha<1$. The two branches therefore separate rather than converge,
\beg\label{nogo}
\Omega^2_{\rm edge}(q)-\Omega^2_{BS}(q)
=\big[4\Delta^2-\Omega^2_{BS}(0)\big]+(1-\alpha)\,(v_Fq)^2 
\en
and the BS mode sinks progressively deeper beneath the edge: a bound
state cannot overtake the threshold it is bound beneath. Both terms on
the right-hand side are positive, so the separation grows monotonically
with momentum. There is consequently no
degeneracy, no level repulsion, and no splitting to measure.
Figure~\ref{Fig1-Schematic} shows this explicitly: over the momentum range
studied the separation grows from $0.18$ to $0.47$ in units of
$2\Delta$.

What this coupling does produce in a clean superconductor is a
redistribution of spectral weight. The BS mode acquires amplitude-channel
character in proportion to $\Pi^{\times}_{sd}$, visible in the
off-diagonal weight $A_{14}(q,\Omega)$, which is nonzero only through the
particle--hole asymmetric coupling: it vanishes identically at $\bq=0$, and
again when $\mathbf q$ is rotated to the nodal direction
$\phi_{\mathbf q}=\pi/4$. Because $A_{14}(q,
\Omega)$ is generated solely by the particle--hole asymmetric
coupling, it vanishes identically in its absence. In other words,  an experiment
searching for it looks for a signal against zero rather than for a small
change in a large background. The effect is thus not resonantly enhanced but unambiguous. We find
in addition that the BS mode remains a genuinely sharp sub-gap pole at
every momentum: for an isotropic $s$-wave condensate the two-quasiparticle
continuum is bounded below by $2\Delta$ at all $\bq$ --- the $d$-wave form
factor is a property of the pairing vertex, not of the gap, and cannot open
decay phase space below $2\Delta$ --- so the bound state has nothing to
decay into, and its width vanishes identically in the clean limit at $T=0$
(thermal quasiparticles restore only an exponentially small width
$\propto e^{-\Delta/T}$).

The kinematic obstruction discussed above, however, is specific to the clean limit, and
identifying where it fails points to the regime in which a genuine
Higgs--BS resonance should be sought. What must change is the pinning of
the amplitude resonance to the pair-breaking threshold, since it is this
pinning that forces the amplitude branch to inherit the threshold
dispersion with coefficient unity. Recently, Nosov, Andriyakhina and Burmistrov~\cite{Nosov2025} have shown that disorder
detaches the Schmid--Higgs resonance from the pair-breaking edge. The
mode remains a resonance --- a pole of the pair susceptibility
analytically continued through the branch cut of the two-quasiparticle
continuum onto the second Riemann sheet --- but its position is no
longer locked to the threshold: for $\tau\Delta\ll1$ it lies below
$2\Delta$ and disperses {downward},
\beg\label{wSHdirty}
\Omega_{SH}(q)\simeq\Delta\left[2-\frac{4\xi^4q^4}{\pi^2}
\ln^2\!\left(\frac{2\sqrt{\pi}}{\xi q}\right)\right],
\en
crossing over to the clean behaviour
$\Omega_{SH}(q)\simeq\Delta(2+\xi^2_{\rm clean}q^2/8)$ only as
$\tau\Delta\to\infty$ (here $\tau^{-1}$ denotes the scattering rate due to potential impurities). Note that this clean-limit form describes the second-sheet pole, whose coefficient (1/2 in two dimensions) originates from the Fermi-surface average of $({\mathbf v}_F\bq)^2$, whereas the unit coefficient in Eq. \eqref{edge} is set by the extremal antinodal pair configuration that defines the threshold. The two are compatible: as $\tau\Delta\to\infty$ the pole retreats beneath the branch point and ceases to shape the on-axis response, whose maximum rides the edge — and it is the edge dispersion, not the pole dispersion, that enters the kinematic argument of Eq. \eqref{nogo}.

Between these two limits the amplitude dispersion
must pass through zero, so that at moderate disorder,
$\tau\Delta\sim1$, the amplitude branch is nearly flat --- precisely the
configuration that the intuitive picture of a level pinned at $2\Delta$
tacitly assumes, and which the clean theory does not in fact provide.
Writing the amplitude branch as
$\Omega^2_{SH}(q)=\Omega^2_{SH}(0)+\alpha_{SH}(\tau\Delta)(v_Fq)^2$, a crossing with the BS branch requires only
$\alpha_{SH}(\tau\Delta)<\alpha_{BS}$, i.e.
\beg\label{qstar}
(v_Fq^*)^2=\frac{\Omega^2_{SH}(0)-\Omega^2_{BS}(0)}
{\alpha_{BS}-\alpha_{SH}},
\en
which is guaranteed once $\alpha_{SH}$ softens through zero while the BS
continues to disperse upward; in the clean limit
$\alpha_{SH}>\alpha_{BS}$ and Eq.~\eqref{qstar} has no solution, which
is the content of the no-go above. Two effects push in the same
direction as disorder is increased: the amplitude branch flattens, and,
disorder being pair-breaking for the anisotropic subdominant channel \cite{Hirschfeld2011}, it
simultaneously weakens $\lambda_d$ and raises $\Omega_{BS}(0)$ toward the
edge, reducing the numerator of Eq.~\eqref{qstar}. The splitting at the
resulting crossing is then set by the coupling computed in the present
work, $2|\Pi^{\times}_{sd}(q^*)|$. This scenario is sketched in
Fig.~\ref{fig:disorder}.

Our paper is organized as follows. Section~II introduces the model Hamiltonian. Section~III contains all the necessary technical details of our study where we derive the coupled
susceptibility matrix and the master formula for the cross-bubble. Section~IV contains the analysis of the collective modes: the branch dispersions, the channel-resolved spectral functions, and the kinematic obstruction to a resonant Higgs--BS hybridization.
Discussion of our results along with experimental estimates is given in Section~V. Finally, in Section VI we summarize the main results of this work. All relevant technical details can be found in Appendices. 

\section{Model}
We consider a one band model of interacting spin-$1/2$ fermions in two spatial dimensions. We assume that there are two interaction Cooper channels 
with $s$-wave and $d-$wave symmetries 
\beg\label{Eq1}
\begin{aligned}
&\hat{\cal H}=\sum_{\bk,\sigma} \xi_\bk \hat{c}^\dagger_{\bk,\sigma}\hat{c}_{\bk,\sigma}\\&+
    \sum_{\bk\bp\bq}\sum\limits_{\mathrm{a}=s,d}V_{\mathrm{a}}(\bk,\bp) \hat{c}^\dagger_{\bk+\frac{\bq}{2},\uparrow} \hat{c}^\dagger_{-\bk+\frac{\bq}{2} ,\downarrow}\hat{c}_{-\bp+\frac{\bq}{2} ,\downarrow}\hat{c}_{\bp+\frac{\bq}{2},\uparrow},
\end{aligned}
\en
 where 
 $\hat{c}^\dagger$ ($\hat{c}$) are the creation (annihilation) fermionic operators,
 $V_{\mathrm{a}}$ is the pairing interaction in $s$-wave (leading) and $d$-wave (subleading) channels, and 
 $\xi_\bk=k^2/(2m) - \veps_F$ and $\veps_F$ is the Fermi energy. 
     We approximate $V_{\mathrm{a}}(\bk,\bp)$ as  
\begin{align}
    V_{\mathrm{a}}(\bk,\bp)=- \left(\frac{\lambda_{\mathrm{a}}}{\nu_F}\right){\gamma}_{\mathrm{a}}(\theta_\bk)\, {\gamma}_{\mathrm{a}}(\theta_\bp),
   \label{Pairing Interaction in s-d}
\end{align}
 where $\nu_F = m/(2\pi)$ is the density of states at the Fermi level 
  $\lambda_{s,d}>0$ are the dimensionless coupling constants for the corresponding channels, $\gamma_s(\theta_\bk)=1$ and $\gamma_d(\bk)=\sqrt{2}\cos2\theta_\bk$ are the normalized $s$- and $d$-wave form factors ($\theta_\bk$ defines the direction of the momentum on the Fermi surface 
  with respect to $\hat x$:  $\bk_F=k_F(\cos\theta_\bk,\sin\theta_\bk)$). As we have mentioned above it is assumed that $\lambda_s>\lambda_d$.

In addition 
to  $V_{s,d}$, 
 we include 
 the long-range Coulomb interaction between fermions
\begin{align}
    \hat{H}_{C} = \dfrac{1}{2}\sum_\bq V_\bq \rho_\bq \rho_{-\bq},
\end{align}
where $\rho_\bq=\sum_{\bk,\sigma} c^\dagger_{\bk+\bq,\sigma}c_{\bk,\sigma}$ is the particle density operator and  $V_\bq$ is the Coulomb potential $V_\bq=2\pi e^2/|\bq|$. Thus, the full model Hamiltonian is $\hat{H} =   \hat{H}_{C} + \hat{\cal H}$.

  At $T=0$, Eq.~\eqref{Eq1} describes a superconductor with a $s-$wave pairing gap
  $\Delta_\bk=\Delta$. The gap magnitude is obtained by solving the non-linear 
  BCS-like gap equation
  \begin{align}
     1=\frac{\lambda_s}{2}\int\limits_{-\Lambda}^\Lambda\dfrac{d\xi_\bk}{ \sqrt{\xi^2_\bk+\Delta_\bk^2}},
     \label{Gap equation s-d}
     \end{align}
where $\Lambda$ is an ultraviolet cutoff,
 which we introduce symmetrically with respect to $\xi_\bk$. 

\section{Pair susceptibility matrix}
Theoretical analysis of the collective modes which are governed by our model Hamiltonian can be performed using diagrammatic \cite{AGD} or quasiclassical approaches \cite{Eilenberger1968, Larkin1965,LarkinVertex,ASchmid1968,SereneRainer1983, BELZIG1999,Kopnin2001}. In the context of studies of collective modes, it has been shown that both approaches yield the same results \cite{Kazi2026}. 
In this paper, we will also use both of these approaches. The quasiclassical approach is based on the following equation
\beg\label{EilenMain}
[\eps\check{\tau}_3+\check\Delta_\bn(\br,t)\stackrel{\circ},\check{g}]+\frac{i}{2}\left\{\check{\tau}_3,\partial_t\check{g}\right\}+i{v}_F(\bn\mbox{\boldmath $\nabla$}_\br)\check{g}=0,
\en
where $v_F$ is the Fermi velocity,  $\check{g}$ is the quasiclassical propagator defined in Keldysh and Nambu spaces
\beg\label{KeldyshProps}
\check{g}=\left(\begin{matrix} \hat{g}^R & \hat{g}^K \\ 0 & \hat{g}^A\end{matrix}\right).
\en
and the convolution should be understood as follows
\beg\label{SecondTerm}
\begin{split}
&\check{A}{\circ}\check{B}=\check{A}_{\bn\eps}(\br,t)e^{\frac{i}{2}(\stackrel{\leftarrow}\partial_\eps\stackrel{\rightarrow}\partial_t-\stackrel{\leftarrow}\partial_t\stackrel{\rightarrow}\partial_\eps)}
\check{B}_{\bn\eps}(\br,t)
\end{split}
\en
Here we are ignoring the gradients with respect to the spatial coordinate and momentum. The order parameter is given by sum of its components in the longitudinal (amplitude) and transverse (phase) channels 
\beg\label{hatDelta}
\hat{\Delta}=i\hat{\tau}_2\Delta^{(1)}+i\hat{\tau}_1\Delta^{(2)}
\en
and each of these two components can be found from the self-consistency equation
\beg\label{Self}
\Delta^{({\overline{a}})}=\frac{\lambda_{\textrm{a}}}{2}\int\limits_{-\infty}^\infty{d\eps}\textrm{Tr}\left\{-i\hat{\tau}_{a}\hat{g}^K\right\}
\en
and $\overline{1}=2$.

Quasiclassical propagator $\check{g}$ is subject to a normalization condition
\beg\label{norm}
\check{g}\circ\check{g}=\check{{\mathbbm{1}}}.
\en
In equilibrium we set $\hat{g}_{\eps}^{R(A)}=\hat{\tau}_3 g_{\eps}^{R(A)}+i\hat{\tau}_2f_{\eps}^{R(A)}$, 
$\hat{\Delta}=i\hat{\tau}_2\Delta$, $g_{\eps}^R={\eps}/{\eta_{\eps}^R}$,  
$f_{\eps}^R={\Delta}/{\eta_{\eps}^R}$ and 
functions $\eta_{\bn\eps}^{R(A)}$ are given by 
\beg\label{etaRA}
\eta_{\eps}^{R(A)}=\left\{\begin{split} &\pm\mathrm{sign}(\eps)\sqrt{(\eps\pm i\delta)^2-\Delta^2}, \quad |\eps|\geq\Delta, \\
&i\sqrt{\Delta^2-\eps^2}, \quad |\eps|<\Delta.
\end{split}
\right.
\en
The advanced component of $\check{g}$ can be found using the rule $\hat{g}^A=-\hat{\tau}_3\left[\hat{g}^R\right]^\dagger\hat{\tau}_3$. Finally, in the ground state $\hat{g}^K=(\hat{g}^R-\hat{g}^A)\tanh(\eps/2T)$ is a parametrization enforced by \eqref{norm} \cite{SereneRainer1983,Kopnin2001,Kamenev2009,Kamenev2011}.

\subsection{Longitudinal order parameter fluctuations}
Let us now consider the longitudinal order-parameter fluctuations in both the leading and subleading pairing channels. Consequently, we generalize \eqref{Self} and promote $\delta\Delta^L$ describing the longitudinal fluctuation to a vector in channel space,
\begin{equation}\label{deltaDeltaL}
\delta\hat{\Delta}_{\bn}^L(\bk,\omega)=\sum\limits_{\textrm{a}=s,d}(-i\hat{\tau}_2)\,\delta\Delta^{L}_a(\bk,\omega)\,{\gamma}_{\mathrm{a}}(\theta_\bn).
\end{equation}
At linear order, the $i\hat{\tau}_2$ (amplitude) and $i\hat{\tau}_1$ (phase) sectors stay
decoupled unless particle–hole symmetry is broken by the intrinsic $O(\Delta/\varepsilon_F)$
band asymmetry, or externally by a drive or supercurrent. Therefore, fluctuations in both of these
channels can be treated separately. 

Given \eqref{deltaDeltaL} self-consistency equation generalizes to
\beg\label{Selfa}
\begin{aligned}
\delta\Delta_{a}^L(\bk,\omega)&=\frac{\lambda_a}{2}\int\limits_0^{2\pi}\frac{d\theta_\bn}{2\pi}{\gamma}_a(\theta_\bn)\\&\times\int\limits_{-\infty}^\infty{d\eps}\textrm{Tr}\left\{-i\hat{\tau}_2\delta\hat{g}^K(\bn\eps;\bk\omega)\right\}.
\end{aligned}
\en
Here $\delta\hat{g}^K(\bn\eps;\bk\omega)$ represents a perturbative solution of the Eilenberger equation \eqref{EilenMain} to the leading order in $\delta\hat{\Delta}$ (see Appendix \ref{AppendixA} and Refs. \cite{LiDzero2024,Kamenev2025,Aidan2026} for details).

\begin{figure}[t]
\includegraphics[width=0.9\linewidth]{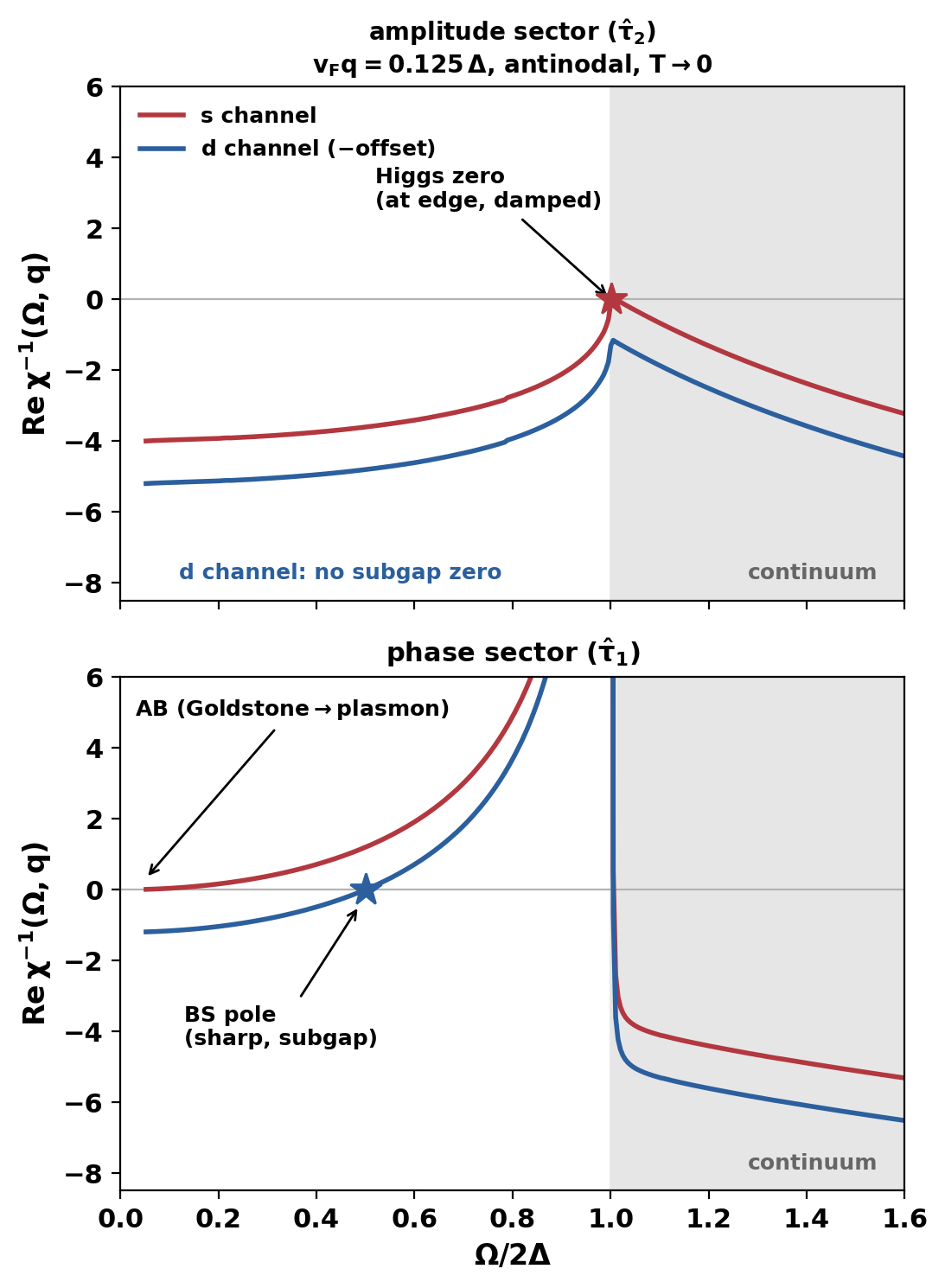}
\caption{\footnotesize Collective-mode classification of an $s$-wave superconductor with a
subdominant $d$-wave channel, from the computed inverse pair
susceptibilities at $v_Fq=0.125\,\Delta$ (antinodal propagation, $T\to0$;
the $d$-channel curves are shifted by the offset
$1/\lambda_d-1/\lambda_s=1.2$). Shown are the diagonal elements of each
sector's $2\times2$ channel matrix. The intra-sector coupling
$\propto(v_Fq)^2\cos2\phi_{\mathbf q}$ shifts the displayed zeros
only at $O((v_Fq)^4)$, invisible at this momentum, while the
cross-sector elements vanish identically in the particle-hole symmetric
theory. Top: amplitude ($\hat\tau_2$) sector. The
$s$-channel zero---the Higgs mode---sits at the pair-breaking edge, where
it is degenerate with the two-quasiparticle continuum (shaded); the
$d$-channel function has no subgap zero, i.e., the amplitude sector hosts
no sharp subdominant mode. Bottom: phase ($\hat\tau_1$) sector. The
$s$-channel zero at $\Omega\to v_Fq/\sqrt2$ is the Anderson--Bogoliubov
mode, promoted to the plasmon by the Coulomb interaction; the $d$-channel
zero below the edge is the sharp BS exciton. The two sharp sub-edge
objects of the problem---Higgs and BS---thus reside in different
particle-hole sectors, which decouple at all momenta in a particle-hole
symmetric theory: their hybridization requires the $O(\Delta/\veps_F)$
coupling computed in this work.}
\label{fig:taxonomy}
\end{figure}

\begin{figure}[t]
\includegraphics[width=0.995\columnwidth]{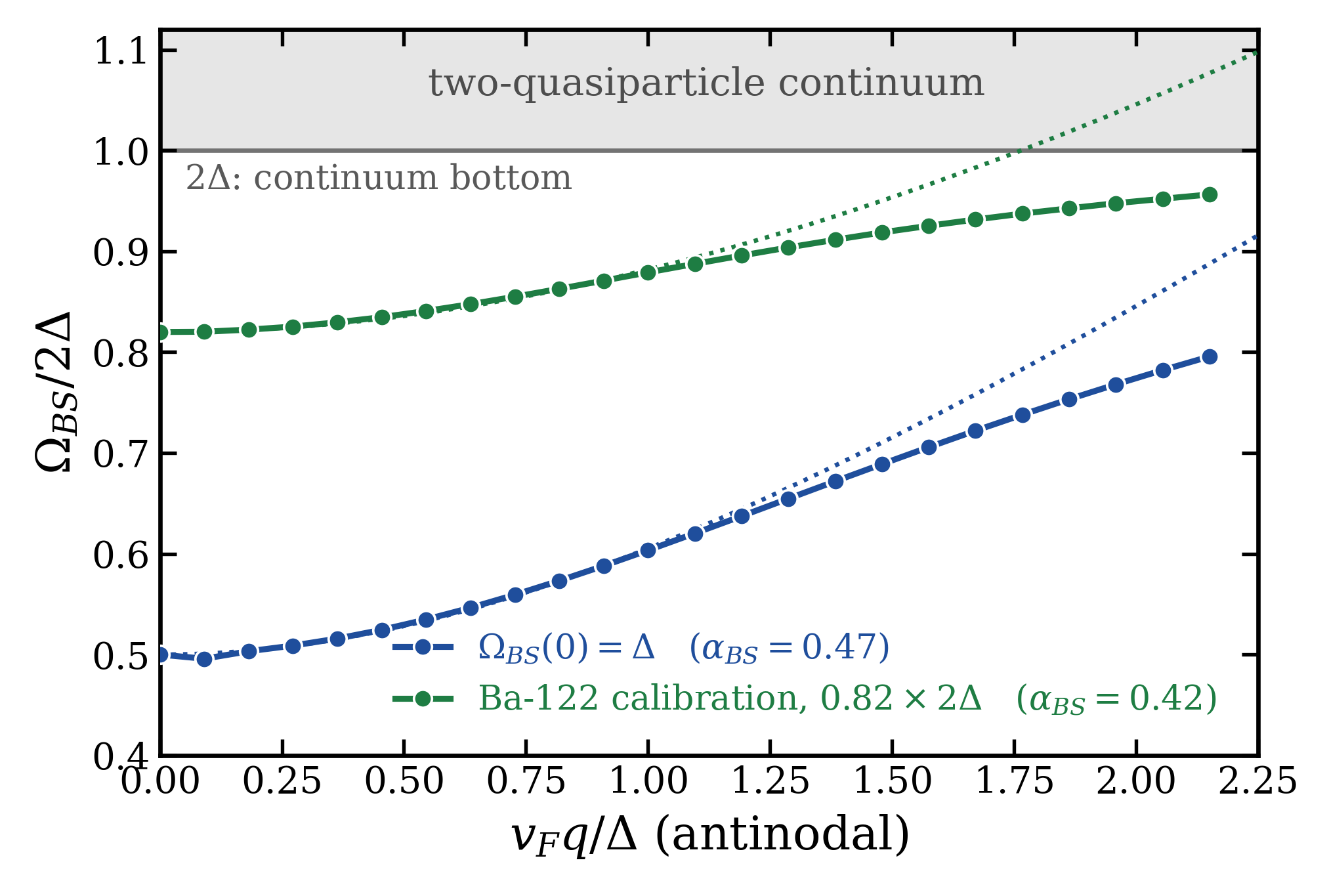}
\caption{\footnotesize Dispersion of the bare Bardasis-Schrieffer (BS)
mode along the antinodal direction, obtained from the zeros of the computed
phase-sector susceptibility for two calibrations of the subdominant
coupling: $\Omega_{BS}(0)=\Delta$ (offset $1/\lambda_d-1/\lambda_s=1.20$;
blue) and the Ba$_{0.6}$K$_{0.4}$Fe$_2$As$_2$ calibration
$\Omega_{BS}(0)=0.82\times2\Delta$ (offset $5.46$, fixed by the Raman mode
energy of Refs.~\cite{Kretzschmar2013,Bohm2014}; green). Dotted curves show
the small-$q$ law $\Omega^2=\Omega_0^2+\alpha_{BS}(v_Fq)^2$ fitted on
$v_Fq\le\Delta$, with the coefficient $\alpha_{BS}=0.47$ ($0.42$) for the blue
(green) branch. At larger momenta the branches fall below the quadratic law
as they are repelled by the flat bottom of the two-quasiparticle continuum
at $2\Delta$, which they approach without crossing: the level repulsion
that prevents the bare BS branch from ever reaching the Higgs
(Sec.~\ref{Anticrossing}).}
\label{fig:bsdisp}
\end{figure}
Thus we obtain a $2\times 2$ system of linear equations which describes hybridization between the longitudinal order parameter fluctuations in the leading and subleading channels: 
\beg\label{SH2x2}
\left(
\begin{matrix} \Pi_{ss}^{||}& \Pi_{sd}^{||} \\
\Pi_{sd}^{||} & \Pi_{dd}^{||} \end{matrix}
\right)\left(\begin{matrix} \delta\Delta_s^L \\ \delta\Delta_d^L\end{matrix}\right)=0,
\en
where we adopted the notations $\Pi_{ab}^{||}=[\chi_{\textrm{SH}}^{-1}]_{ab}$.
Since the $s$-wave channel has been assumed to be a leading one, we have
\beg\label{Couplings}
\frac{1}{\lambda_d}-\frac{1}{\lambda_s}=\delta\lambda_{sd}>0.
\en
In the top panel of Fig. \ref{fig:taxonomy} we show the frequency dependence of the real part of the diagonal elements of matrix \eqref{SH2x2} with the $d$-wave part offset by $\delta\lambda_{sd}$. Thus, the amplitude block which describes the Schmid-Higgs mode has been built. Now we need to build the phase block which contains the physics of Bardasis-Schrieffer mode. 

\subsection{Transverse order parameter fluctuations}
We now consider the case when the system has been subjected to an external perturbation which in turn produces a change in the order parameter corresponding to the fluctuation in the transverse (phase) channel. In full analogy with \eqref{deltaDeltaL} we will describe this by a function
\begin{equation}\label{deltaDeltaT}
\delta\hat{\Delta}_{\bn}^T(\bk,\omega)=\sum\limits_{\textrm{a}=s,d}(-i\hat{\tau}_1)\,\delta\Delta^{T}_a(\bk,\omega)\,{\gamma}_{\mathrm{a}}(\theta_\bn).
\end{equation}
The first Pauli matrix ensures that the resulting corrections to $\check{g}$ do indeed describe the excitation of the phase mode. For instance, the normal part of $\check{g}$ will be proportional to $\hat{\tau}_0$, which will allow us to compute the change in the particle density due to the phase fluctuations of the order parameter in the transverse channel.  Indeed, 
as it is well known, an excitation of the phase mode necessarily produces redistribution of electronic charge. This process is accounted for by the corresponding variation of the Coulomb potential, which is described by a function 
\beg\label{dPhi}
\delta\Phi(\br,t)=\delta\Phi_{\bk\omega}e^{2i(\bk\br-\omega t)}.
\en
With these provisions, using the quasiclassical equation \eqref{EilenMain} it can be shown that the second-order corrections for each element of the matrix $\check{g}$ can be found by solving the following equation
\beg\label{Eq4g2RAT}
\begin{aligned}
&\left[\eps\hat{\tau}_3+\hat{\Delta}_\bn,\delta\hat{g}\right]-2{v}_F(\bn\cdot\bk)\delta\hat{g}+\omega\left\{\hat{\tau}_3,\delta\hat{g}\right\}\\&=\delta\hat{\Delta}_\bn^T\hat{g}_{\bn\eps-\omega}-\hat{g}_{\bn\eps+\omega}\delta\hat{\Delta}_\bn^T\\&+\delta\Phi_{\bk\omega}\hat{g}_{\bn\eps-\omega}-\hat{g}_{\bn\eps+\omega}\delta\Phi_{\bk\omega}.
\end{aligned}
\en
Equation \eqref{Eq4g2RAT} is supplemented by the self-consistency condition for $\delta\Delta^{T}_a(\bk,\omega)$ (see Appendix \ref{AppendixA}) and the Poisson equation
\beg\label{Poisson}
\begin{aligned}
\left(-\frac{k}{\pi e^2}\right)\delta\Phi_{\bk \omega}&=2\nu_F\delta\Phi_{\bk\omega}\\&+\frac{\nu_F}{4}\int\limits_0^{2\pi}\frac{d\theta_\bn}{2\pi}\int\limits_{-\infty}^\infty{d\eps}\textrm{Tr}\left\{\delta\hat{g}^K\right\}.
\end{aligned}
\en
The expression appearing on the left-hand side is equivalent to 
$-V_C^{-1}(\bk)\delta\Phi_{\bk \omega}/\nu_F$ with the 2D Coulomb kernel $V_C(\bk) = 2\pi e^2/k$  evaluated at the potential's own wavevector $2\bk$, Eq. \eqref{dPhi}. Since the quasiclassical approach only takes into account the polarizability of the states near the Fermi surface, the first term on the right-hand side complements it by taking into account the polarizability of the underlying Fermi sea with compressibility given by $\partial n/\partial\mu=2\nu_F$. Solving \eqref{Eq4g2RAT} and inserting the solution into the self-consistency and Poisson equations yields the following system of linear equations:
\beg\label{MyTSystem}
\begin{aligned}
&\chi_{\textrm{AB}}^{-1}(\bk,\omega)\cdot\delta\Delta_{\bk\omega}^T+i\rho({\bk,\omega})\cdot\delta\Phi_{\bk\omega}=0, \\
&\left[4+\chi_{\textrm{CG}}^{-1}(\bk,\omega)+\frac{2k}{\pi\nu_Fe^2}\right]\delta\Phi_{\bk\omega}=i\rho({\bk,\omega})\delta\Delta_{\bk\omega}^T.
\end{aligned}
\en
Functions $\chi_{\textrm{AB}}^{-1}(\bk,\omega)$, $\chi_{\textrm{CG}}^{-1}(\bk,\omega)$ and $\rho(\bk,\omega)$ have been defined in Appendix \ref{AppendixA}. It is straightforward to check that for the case of small momentum \eqref{MyTSystem} gives gapless plasmon mode $\omega_{\textrm{pl}}(k)=\sqrt{2\pi e^2\nu_Fv_F^2k}$ \cite{Kazi2026}.

In the presence of the sub-leading pairing channel system \eqref{MyTSystem} generalizes to 
\beg\label{LargeTSystem}
\left(
\begin{matrix} \Pi_{ss}^{\perp} & \Pi_{sd}^{\perp} & i\rho_s \\
\Pi_{sd}^{\perp} & \Pi_{dd}^{\perp} & i\rho_d \\ -i\rho_s & -i\rho_d & D_\Phi\end{matrix}
\right)\left(\begin{matrix} \delta\Delta_s^T \\ \delta\Delta_d^T \\ \delta \Phi\end{matrix}\right)=0,
\en
where similar to \eqref{SH2x2} we use $\Pi_{ab}^{\perp}=[\chi_{\textrm{AB}}^{-1}]_{ab}$ and ${\cal D}_\Phi(\bk,\omega)=4+\chi_{\textrm{CG}}^{-1}(\bk,\omega)+2k/(\pi\nu_Fe^2)$. In \eqref{LargeTSystem} $[\chi_{\textrm{AB}}^{-1}]_{\textrm{ab}}$ and $\rho_{\mathrm{a}}$ are evaluated with the ground state propagators but with the angular integrals weighted by the corresponding pairing form factors.  The diagonal element $\chi_{\textrm{AB},dd}^{-1} $ carries the physics of the Bardasis-Schrieffer mode, Fig. \ref{fig:bsdisp}, while $\chi_{\textrm{AB},ss}^{-1} $ describes the emergence of the plasmon mode (see bottom panel in Fig. \ref{fig:taxonomy}). Notably, the matrix element $\rho_d$ describes the hybridization between the BS mode and the plasmon mode discussed in \cite{SunMillis2020} (see below).

\subsection{Hybridization between the amplitude and phase modes}\label{Anticrossing}
As we have discussed in the introduction, hybridization between longitudinal and transverse fluctuations in the leading and sub-leading channels is not generally possible even at finite momentum, for, as we will explicitly demonstrate below, it also requires particle-hole asymmetry.
Intuitively, this can be understood as follows. An amplitude oscillation of the order parameter $\Delta(\br,t)=|\Delta(\br,t)|e^{i\varphi(\br,t)}$ modulates the gap magnitude $|\Delta(\br,t)|$. In a particle-hole symmetric case, the amplitude
variations produce equal and opposite charge on particle-like and hole-like
branches and so the net charge variation vanishes exactly. In this case the amplitude mode never hybridizes with the phase sector and therefore remains
charge-neutral, i.e. it redistributes spectral weight
\emph{symmetrically} about the Fermi level. Fluctuation of a phase, on the other
hand, carries charge --- by the Josephson relation $\partial_t\varphi$ is a
local chemical-potential (charge-imbalance) shift, and a phase gradient
${\vec \nabla}_\br\varphi$ is a supercurrent. Coupling the amplitude and phase modes therefore requires
converting a charge-neutral motion into a charge-carrying one. But charge is
precisely the imbalance between particles and holes, so this conversion is
possible only if the particle and hole sides of the Fermi sea are inequivalent,
i.e.\ only if the particle-hole symmetry is broken. It was recently argued that this mechanism allows one to probe the Schmid-Higgs resonance 
in the $dc$-response of the unconventional superconductor \cite{Kozii2026}.

Since the quasiclassical equation \eqref{EilenMain} is by construction particle-hole symmetric, there are two options at our disposal. The first one consists of introducing an electrochemical potential term into \eqref{EilenMain} and retaining linear momentum gradients \cite{Kozii2026}. These terms produce corrections to the response functions $O(\Delta/\veps_F)$ ($\veps_F$ is the Fermi energy). Another option is to use a diagrammatic approach \cite{Kazi2026} and compute the corresponding contribution to the pair susceptibility matrix in the Matsubara representation and then perform the analytic continuation to real frequencies. This will allow us to explicitly account for the hybridization between the amplitude and phase modes. As it has been shown in Ref. \cite{Kazi2026} both approaches lead to identical results, so here we choose the second ("diagrammatic") approach and consider the following response function:
\beg\label{PisdCross}
\begin{aligned}
\Pi_{\textrm{ab}}^{\times}(\bq,i\Omega_l)&=\frac{T}{2}\sum\limits_{i\omega_n}\int\frac{d^2\bk}{(2\pi)^2}{\gamma}_{\textrm{a}}(\theta_\bk){\gamma}_{\textrm{b}}(\theta_\bk)\\&\times\textrm{Tr}\left\{\hat{\tau}_2\hat{G}\left(\frac{\bq}{2}+\bk,\frac{z}{2}+i\omega_n\right)\right.\\&\times\left.\hat{\tau}_1\hat{G}\left(\frac{\bq}{2}-\bk,\frac{z}{2}-i\omega_n\right)\right\},
\end{aligned}
\en
where $\hat{G}^{-1}(\bk,i\omega_n)=i\omega_n\hat{\tau}_0-\xi_\bk\hat{\tau}_3-\hat{\tau}_1\Delta$, $\omega_n=\pi T(2n+1)$ is a fermionic Matsubara frequency and $z=\Omega_l=2\pi Tl$ is the bosonic Matsubara frequency. Calculation of the trace followed by the summation over the Matsubara frequencies yields
\beg\label{SummedUp}
\begin{aligned}
&\Pi_{\textrm{ab}}^{\times}(\bq,i\Omega_l)=\frac{\Omega_l}{2}\!\int\!\frac{d^2\bk}{(2\pi)^2}\,
{\gamma}_{\textrm{a}}(\theta_\bk){\gamma}_{\textrm{b}}(\theta_\bk)\\&\times
\left[
\frac{(\frac{\xi_{+}}{E_{+}}-\frac{\xi_{-}}{E_{-}})[n_F(E_+)-n_F(E_-)]}{(i\Omega_l)^2-(E_+-E_-)^2}\right.\\&\left.
-\frac{(\frac{\xi_{+}}{E_{+}}+\frac{\xi_{-}}{E_{-}})[1-n_F(E_+)-n_F(E_-)]}{(i\Omega_l)^2-(E_++E_-)^2}
\right].
\end{aligned}
\en
In this expression $E_{\pm}=E_{\frac{\bq}{2}\pm\bk}$, $\xi_{\pm}=\xi_{\frac{\bq}{2}\pm\bk}$ and $n_F(\veps)$ is the Fermi distribution function. The physical meaning of the two contributions under the integral is quite clear: the first term is a quasiparticle term which contributes at $T\not =0$ and accounts for Landau damping, while the second term is a pair-breaking term and it becomes dominant when $E_{+}+E_{-}\geq 2\Delta$.

In order to make further progress, we will analyze \eqref{SummedUp} in the limit $T\ll \Delta$ and at small values of momentum $\bq$. In a low-$T$ limit, $n_F(E_{\frac{\bq}{2}\pm \bk})\to 0$, so that the first term under the integral vanishes identically. We approximate $\xi_{\frac{\bq}{2}\pm\bk}\approx\xi_\bk\pm v(\xi)(\bn\bq)/2+O(q^2)$, where $v(\xi)=v_F\sqrt{1+\xi/\veps_F}$. Note that the single-particle density of states is constant in $2D$, and as a consequence $v(\xi)$ is the only source of the particle-hole asymmetry. 

Expanding the expression under the integral for small $\bq$ and evaluating the remaining integrals up to the leading order in $v_Fq/\veps_F$ we find:
\begin{equation}
\Pi_{\textrm{ab}}^{\times}(\bq,i\Omega_l)\approx-\,\frac{\nu_F\Omega_l}{4\veps_F}\left(\frac{v_Fq}{\Delta}\right)^2G_{\mathrm{ab}}\left(\theta_\bq,\frac{i\Omega_l}{2\Delta}\right),
\label{eq:final}
\end{equation}
where function $G_{\mathrm{ab}}$ has been defined in Appendix \ref{AppendixC}. The dependence of $\Pi_{\textrm{ab}}^{(\times)}(\bq,i\Omega_l)$ on real frequencies is obtained by performing an analytic continuation $i\Omega_l\to \Omega+i0$ which is also described in Appendix \ref{AppendixC}. 

A few comments are in order. The diagonal elements of $\Pi_{\textrm{ab}}^{\times}$ vanish as $q\to 0$: this is the consequence of the fact that we work in two spatial dimensions. In three dimensions there will be an extra contribution to the diagonal terms $\propto \nu_F'\Omega$ which clearly remain finite at $\bq=0$
(no such contribution arises in the off-diagonal parts due to the orthogonality of the pairing form factors). Matrix element $\Pi_{ss}^\times$ describes the coupling between the charge plasmon sector and the amplitude mode, and it reproduces the physics discussed in Ref. \cite{SunMillis2020}.
In this regard, we would like to emphasize that the genuinely new object is the direct off-diagonal Higgs-BS coupling given by $\Pi_{sd}^\times(\bq,\Omega)$, which to the best of our knowledge, has not been discussed before and which provides a direct Higgs--BS coupling 
along the antinodal direction. We thus have collected all the required ingredients to evaluate the dispersion of the hybridized amplitude and Bardasis-Schrieffer modes. 

\section{Analysis of the collective modes}
Collecting all the matrix elements, we arrive at the linear system
$\hat M(\bq,\Omega)\,\psi=0$ for the five-component fluctuation vector
$\psi=(\delta\Delta_s^L,\delta\Delta_d^L,\delta\Delta_s^T,
\delta\Delta_d^T,\delta\Phi)^{T}$, i.e.
\beg\label{LargeHiggsBS}
\left(
\begin{matrix} \Pi_{ss}^{||} & \Pi_{sd}^{||} & \Pi_{ss}^{\times} & \Pi_{sd}^{\times} & 0 \\
\Pi_{sd}^{||} & \Pi_{dd}^{||} & \Pi_{sd}^{\times} & \Pi_{dd}^{\times} & 0 \\
-\Pi_{ss}^\times & -\Pi_{sd}^\times & \Pi_{ss}^{\perp} & \Pi_{sd}^{\perp} & i\rho_s \\
-\Pi_{sd}^\times & -\Pi_{dd}^\times & \Pi_{sd}^{\perp} & \Pi_{dd}^{\perp} &  i\rho_d \\  0 & 0 & -i\rho_s & -i\rho_d & {\cal D}_\Phi\end{matrix}
\right)\left(\begin{matrix} \delta\Delta_s^L \\ \delta\Delta_d^L \\ \delta\Delta_s^T \\ \delta\Delta_d^T \\ \delta \Phi\end{matrix}\right)=0.
\en
In Appendix \ref{AppendixC} we show that this system can be further simplified by excluding the Coulomb field and also neglecting terms beyond linear order in $\Delta/\veps_F$. We note that Eq. \eqref{LargeHiggsBS} holds at the particle–hole-symmetric level: the $O(\Delta/\varepsilon_F)$
couplings between the amplitude components and $\delta\Phi$ can appear only in three spatial dimensions as discussed in Appendix \ref{AppendixC}.

The collective modes are the poles of the fluctuation propagator
$\check\chi(\bq,\Omega)=\hat M^{-1}(\bq,\Omega)$, i.e.\ the complex-frequency
zeros of $\det\hat M$. On the real-frequency axis, a mode of finite lifetime
does \emph{not} appear as a zero of $\det\hat M$. It appears as a maximum of
the channel-resolved spectral weight
\beg\label{Aspec}
A_{\mathrm{ab}}(\bq,\Omega)=-\frac{1}{\pi}\,
\mathrm{Im}\,\big[\hat M^{-1}(\bq,\Omega)\big]_{\mathrm{ab}},
\en
where the indices $1,\ldots,5$ label the components of $\psi$ in
\eqref{LargeHiggsBS}. The peak position and half-width of a diagonal weight
$A_{\mathrm{aa}}$ give the corresponding mode frequency and damping; thus
$A_{11}$ is the amplitude ($s$-wave Higgs) weight and $A_{44}$ the
Bardasis--Schrieffer weight, while the off-diagonal $A_{14}$ measures the
amplitude character admixed into the phase sector.
The two sub-edge excitations realize the two cases in opposite ways. The
Higgs mode lies at the $s$-wave pair-breaking edge $\Omega=2\Delta$ and
decays into the two-quasiparticle continuum immediately above it: it is a
damped threshold resonance, and we extract its branch from the maxima of
$A_{11}$ (Higgs, $\delta\Delta_s^{L}$). The Bardasis--Schrieffer mode, by
contrast, is a true bound state. For an isotropic $s$-wave condensate, the
quasiparticle spectrum is fully gapped, and the two-quasiparticle continuum
at total momentum $\bq$ is bounded below by $2\Delta$ for every $\bq$: the
subleading $d$-wave form factor ${\gamma}_d\propto\cos2\theta$ enters
only as a vertex weight in the angular integral of the pair bubble and
cannot open decay phase space below $2\Delta$. Below the continuum bottom
every entry of $\hat M$ is therefore real at $T\to0$, the BS pole ---
calibrated to $\Omega_{BS}(0)=0.82\times2\Delta$ --- is a genuine real
zero of $\det\hat M$, and its intrinsic width vanishes, $\Gamma_{BS}=0$.
We extract the BS branch from $\det\hat M=0$, equivalent to the
$\delta$-function line of $A_{44}$ (BS, $\delta\Delta_d^{T}$).

Before turning to the amplitude--phase hybridization proper, it is worth
locating within Eq.~\eqref{LargeHiggsBS} the finite-momentum physics of
Ref.~\cite{SunMillis2020}, so as to separate it clearly from the coupling
that is the subject of this work. That reference shows that at $\bq\neq0$
the Bardasis--Schrieffer mode acquires a direct coupling to the
charge-density response --- it is dark at $\bq=0$, where the orthogonality
of the pairing form factors forbids it --- and that it undergoes an
avoided crossing with the plasmon, which in a layered geometry disperses
as $\sqrt{q}$ and is therefore low-lying. In the two-dimensional geometry
considered here, the corresponding amplitude--charge elements vanish
identically, since $\nu_F'=0$: the Higgs mode has no direct coupling to
$\delta\Phi$ and reaches the charge sector only through
$\Pi^{\times}_{ss}$ and the $s$-wave phase component.

In the language of Eq.~\eqref{LargeHiggsBS} this is entirely the physics
of the charge-coupling border: the entry $i\rho_d$, which couples the
$\delta\Delta^T_d$ (BS) fluctuation to $\delta\Phi$ and is of order
$(v_Fq)^2\cos2\phi_{\mathbf q}$, is precisely the coupling of the BS mode
to the charge-density response, while the avoided crossing with the
plasmon is generated by this same border acting against the diagonal
$D_\Phi=4+\chi^{-1}_{\textrm{CG}}+2k/(\pi\nu_Fe^2)$, whose zero is the
plasma mode. Crucially, this entire corner of the matrix --- the
couplings $i\rho_s,\,i\rho_d$ and the diagonal $D_\Phi$ --- is present
already in the particle-hole symmetric theory, at order
$O(1)$: it requires only finite momentum, not
particle-hole asymmetry. The BS--plasmon anticrossing of
Ref.~\cite{SunMillis2020} is thus mediated by the charge sector and is
recovered here without further assumptions. 

By contrast, the direct
Higgs--Bardasis--Schrieffer coupling that we compute below resides in the
amplitude--phase block $\Pi^{\times}$, which is antisymmetric, odd in
frequency, and of order $\Delta/\varepsilon_F$: it connects the two
\emph{sharp} sub-gap excitations to each other rather than to the
plasmon, and vanishes in the particle-hole symmetric limit. The two
effects therefore occupy distinct corners of
Eq.~\eqref{LargeHiggsBS} and are parametrically different in origin.

We now turn our discussion to the results of the numerical calculations. We fix $\Delta$ as the unit of
energy, work at $T=4\times10^{-4}\Delta$ (effectively $T=0$), and use a
spectral broadening $\gamma=4\times10^{-5}\Delta$ in the retarded/advanced
propagators, verified to leave the extracted mode positions unchanged under
changes in the values of $\gamma$. The subdominant coupling $\lambda_d$ is fixed by placing the $\bq=0$ BS resonance at the experimental value
$\Omega_{BS}(0)=0.82\times2\Delta$ of
Ba$_{0.6}$K$_{0.4}$Fe$_2$As$_2$~\cite{Kretzschmar2013,Bohm2014}. Moreover, the
$\bq=0$ matrix then block-diagonalizes, with all inter-sector elements
vanishing to $\sim10^{-10}$, which we use as a check of the assembly. We scan $N=24$ momenta in the range $0\le v_Fq/\Delta\le2.15$ along the antinodal
direction $\phi_{\bq}=0$.

\begin{figure}[t]
\includegraphics[width=\columnwidth]{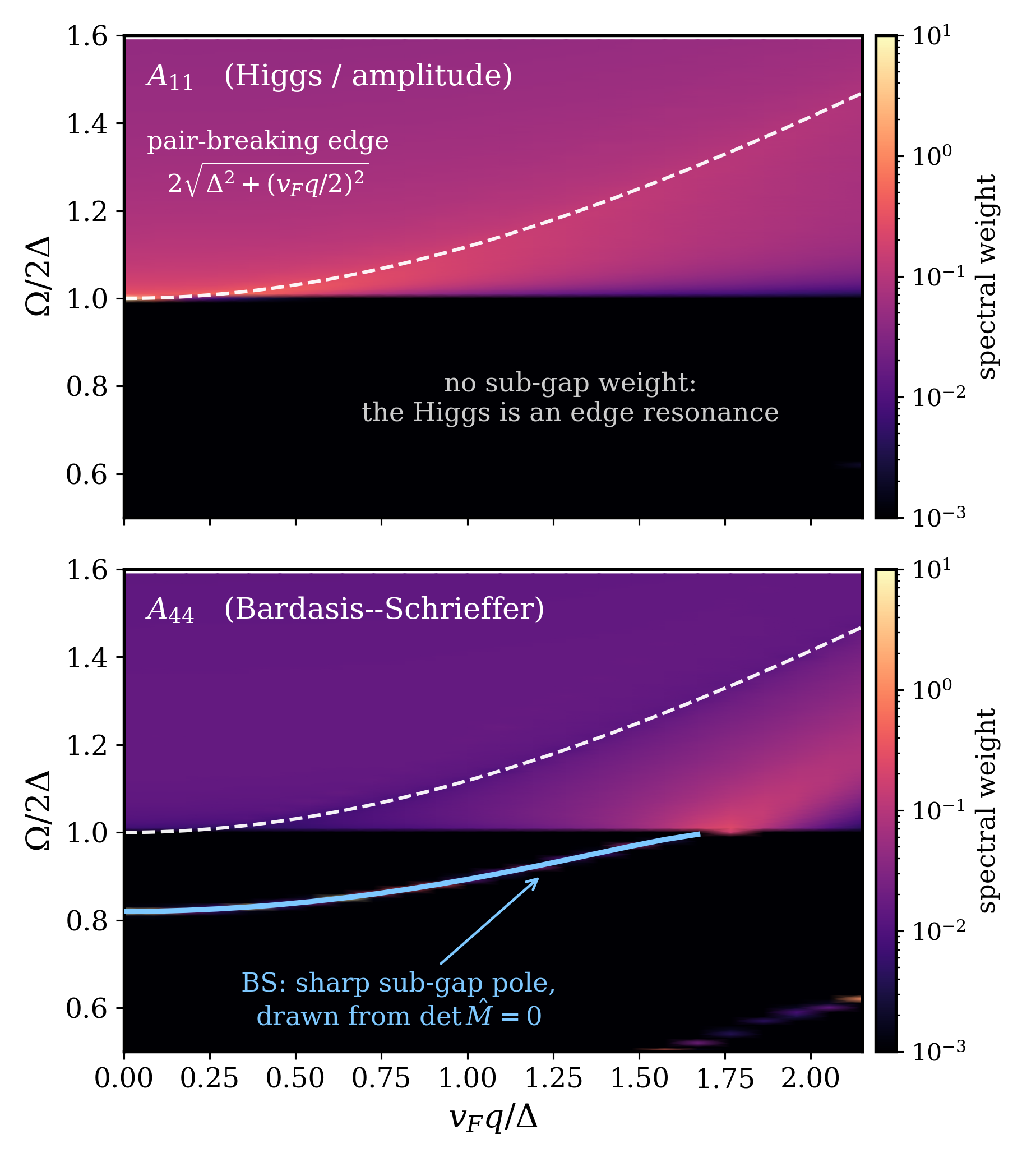}
\caption{\footnotesize Channel-resolved spectral weights, Eq.~\eqref{Aspec},
in the $(\bq,\Omega)$ plane for $\phi_{\bq}=0$, on a logarithmic scale.
Upper panel: the amplitude channel $A_{11}$, which is \emph{identically zero}
below $\Omega=2\Delta$ --- the Higgs possesses no sub-gap pole and exists
only as a threshold resonance riding the pair-breaking edge (dashed). Lower
panel: the Bardasis--Schrieffer channel $A_{44}$. Below $2\Delta$ the
corrected $A_{44}$ is a $\delta$-function line that no finite frequency grid
resolves --- the faint glow beneath the overlay marks grid points that
happen to land on the pole --- so the branch is drawn (solid line) from the
roots of $\det\hat M=0$. Above $2\Delta$ a weak resonance remnant is
visible inside the continuum at the largest momenta. The two branches move
apart with increasing momentum. Both panels show diagonal weights only. The coupling-induced
off-diagonal weight $A_{14}$, of order $\Delta/\veps_F$, lies far
below the color scale of the maps and is not visible here.}
\label{fig:maps}
\end{figure}

In Fig.~\ref{fig:maps} we show the resulting maps. The amplitude channel
carries no spectral weight whatsoever below the pair-breaking threshold:
$A_{11}(\bq,\Omega)$ vanishes for $\Omega<2\Delta$ at every momentum
examined, and its maximum tracks $\Omega_{\mathrm{edge}}(q)$ at every
momentum to within the $0.01\times2\Delta$ resolution of the map.
Independently, the real-axis zeros of $\mathrm{Re}\det\hat M$ above the
edge reproduce $\Omega_{\mathrm{edge}}(q)$ to better than $2\times10^{-4}$
(relative) for $0.27\le v_Fq/\Delta\le1$, a direct machine-precision
confirmation of the unit coefficient in Eq.~\eqref{edge}. This is the numerical content of the statement that the Higgs is a
threshold resonance rather than a bound collective mode: there is no
amplitude pole that could descend to meet the BS.

The BS branch, by contrast, is a genuine sub-gap bound state, and it
disperses upward but slowly. Fitting the pole positions to
$\Omega^2_{BS}(q)=\Omega^2_{BS}(0)+\alpha_{BS}(v_Fq)^2$ on the window
$v_Fq\le\Delta$ gives $\alpha_{BS}\simeq0.50$, consistent with $1/2$ to
our numerical accuracy (the pointwise coefficient varies between $0.494$ and
$0.500$ over $0.27\le v_Fq/\Delta\le1.2$, softening to $\simeq0.46$ at
the largest momenta as the pole is repelled by the flat continuum bottom);
we have not established the value $1/2$ analytically. The precise value is
in any case immaterial, since the inequality $\alpha_{BS}<1$ --- not its
magnitude --- is what forces the branches apart through Eq.~\eqref{nogo}.
The separation between the Higgs and BS branches accordingly \emph{grows}
monotonically, from $0.18$ to $0.47$ in units of $2\Delta$ across the
scanned range, Fig.~\ref{Fig1-Schematic}. Beyond $v_Fq\simeq1.7\Delta$
the binding falls below our frequency resolution,
$1.5\times10^{-3}\times2\Delta$, and the pole becomes numerically
indistinguishable from the continuum bottom, $\Omega_{BS}\to2\Delta^-$.
No crossing occurs, and consequently no avoided crossing.

\section{Discussion}


Our calculation delivers a coupling, but no resonance, and it is worth
separating what is robust in that statement from what is
model-dependent. The absence of an avoided crossing does not rest on the
magnitude of $\Pi^{\times}_{sd}$, nor on the parameters of any
particular material. It follows from Eq.~\eqref{nogo}: the pair-breaking edge disperses with coefficient exactly unity in $(v_Fq)^2$, fixed by
the kinematics of creating two quasiparticles with total momentum
$\bq$, whereas any collective state bound below that continuum
necessarily disperses more slowly, $\alpha<1$. The binding of the BS
mode relative to the threshold therefore \emph{increases} with momentum.

Two features of the numerics deserve comment in this context. First, the
quadratic law holds cleanly only for $v_Fq\lesssim1.2\Delta$: at larger
momenta the bound branch flattens as it is repelled by the flat bottom of
the continuum at $2\Delta$, so the true dispersion is slower than
quadratic and the binding eventually falls below our frequency resolution
($v_Fq\gtrsim1.7\Delta$). Every fitted value satisfies
$\alpha_{BS}<1$, and it is the inequality alone --- not the magnitude ---
that enters Eq.~\eqref{nogo}. Second, the conclusion is insensitive to the
calibration of the subdominant coupling $\lambda_d$: raising
$\Omega_{BS}(0)$ toward the edge reduces the zero-momentum separation
but leaves the $(1-\alpha)(v_Fq)^2$ growth intact, so the branches still
diverge.

The same argument applies verbatim to any sub-gap collective mode of a
competing channel --- Leggett modes of multiband condensates \cite{Leggett1966,Blumberg2007},
mixed-symmetry BS modes --- and shows that finite momentum cannot be
used to tune such a mode into resonance with the Higgs. This is worth
stating explicitly, because the opposite expectation is natural: it
arises whenever the Higgs is idealized as a level pinned at $2\Delta$
rather than as a resonance attached to a dispersing threshold.


The physically observable consequence of the coupling is not a splitting
but an induced amplitude character in the BS mode, controlled by the
off-diagonal spectral weight $A_{14}$, which inherits the full symmetry
structure of $\Pi^{\times}_{sd}$, Eq.~\eqref{PisdCross}. 
Because $A_{14}$ is generated solely by the particle--hole asymmetric
coupling, an experiment searching for it looks for a signal against
zero rather than for a small change on top of a large background. The effect
therefore is not resonantly enhanced, but it is unambiguous, and verifying
the $q^2$ onset together with the $\cos2\phi_{\bq}$ law tests the
selection rule directly.

A second, independent prediction concerns the BS lineshape. In the clean
limit at $T=0$ the BS mode carries no intrinsic width at any momentum:
below $2\Delta$ there is no decay phase space, so the observed linewidth
of the sub-gap peak is set by the instrument (and, at $T>0$, by an
exponentially small thermal contribution $\propto e^{-\Delta/T}$). A
resolution-limited sub-gap line that stays resolution-limited as it
disperses is itself a falsifiable statement: any intrinsic broadening that
grows with $q$ would signal physics beyond the clean two-channel model ---
disorder pair-breaking being the leading candidate, see below. This
sharpness is favorable for the detection of the induced amplitude weight,
which appears at a $\delta$-like line rather than inside a broad hump.


To put these predictions in physical units, we take optimally doped
Ba$_{0.6}$K$_{0.4}$Fe$_2$As$_2$, the one material in which the BS mode
of the subdominant $d_{x^2-y^2}$ channel has been resolved directly:
Raman scattering finds the $B_{1g}$ BS exciton at
$\Omega_{BS}\simeq140~\mathrm{cm}^{-1}=17.4$~meV, below the
pair-breaking edge $2\Delta\simeq170~\mathrm{cm}^{-1}=21.1$~meV of the
electron pockets~\cite{Kretzschmar2013,Bohm2014}, i.e.\
$\Omega_{BS}/2\Delta\simeq0.82$, which is the calibration used
throughout Sec.~IV. With the ARPES Fermi velocity
$\hbar v_F\simeq0.5$~eV\,\AA  \cite{Richard2011}, the unit of momentum is
$\Delta/\hbar v_F\simeq2.1\times10^{-2}$~\AA$^{-1}$, so the full range
scanned in Sec.~IV, $0\le v_Fq/\Delta\le2.15$, corresponds to
$0\le q\le4.5\times10^{-2}$~\AA$^{-1}$ --- squarely inside the window
$q\sim10^{-2}$--$0.5$~\AA$^{-1}$ covered by momentum-resolved electron
energy loss spectroscopy (M-EELS) at few-meV resolution \cite{Vig2017,Husain2019}.

Three quantities are then directly measurable: (i) the BS mode disperses
upward from $17.4$~meV at $\bq=0$ toward the continuum bottom at
$21.1$~meV, with the coefficient $\alpha_{BS}\simeq1/2$; beyond
$q\simeq3.5\times10^{-2}$~\AA$^{-1}$ its binding falls below our
numerical resolution ($\simeq0.03$~meV), so over the scanned range the
peak shifts by $\simeq3.7$~meV; (ii) the line remains sharp throughout:
in the clean limit at low temperature the sub-gap peak has no intrinsic
width, so its observed width is set by the instrument at every momentum;
and (iii) the pair-breaking edge, by contrast, rises from $21.1$ to
$30.9$~meV over the same interval, so the gap between the BS line and the
edge widens from $3.8$ to $9.9$~meV: the divergence of the two features,
and not their approach, is the direct experimental signature of
Eq.~\eqref{nogo}.

Superimposed on this, the induced amplitude weight
$A_{14}\propto(\Delta/\veps_F)(v_Fq)^2\cos2\phi_{\bq}$ grows
quadratically across the range. Its size is set by $\Delta/\veps_F$,
which is not small in the materials of interest: the shallow pockets of
the iron-based superconductors give
$\Delta/\veps_F\sim0.1$--$0.5$~\cite{Kasahara2014}, so that at the upper
end of the momentum range the amplitude admixture is a percent-level
fraction of the BS weight rather than a vanishingly small one. Near-field
terahertz spectroscopy reaches a smaller momentum: a tip of radius
$a_{\rm tip}\simeq10$~nm supplies $q\sim1/a_{\rm tip}\simeq
10^{-2}$~\AA$^{-1}$, i.e.\ $v_Fq\simeq0.5\Delta$, where the $q^2$ law
puts the induced weight some twenty times below its value at the top of
the M-EELS range \cite{Cocker2021}. M-EELS is therefore the natural probe for the effect,
with s-SNOM providing an independent check at small momentum. In either case the decisive test is angular rather than absolute:
rotating $\bq$ from the antinodal to the nodal direction of the
subdominant gap extinguishes $A_{14}$ at fixed instrument settings, so
that a $45^\circ$ rotation of the sample converts the signal into its
own null measurement. 

The obstruction expressed by Eq.~\eqref{nogo} is specific to the clean
limit,  and the estimates that follow (as distinct from the clean-limit numbers just given) 
apply to the moderate-disorder scenario. Fig.~\ref{fig:disorder} shows how the amplitude
resonance has detached from the pair-breaking edge, and its dispersion
has softened to $\alpha_{SH}<\alpha_{BS}$~\cite{Nosov2025} in contrast to
the clean superconductor treated above. We stress that Fig.~\ref{fig:disorder}
is illustrative: establishing it quantitatively requires treating both sectors in the diffusive limit, which lies beyond the clean
quasiclassical framework used here. Nevertheless, we expect that in that regime the crossing condition Eq.~\eqref{qstar} then acquires a
solution for $q^*$, and the coupling computed in this work would set the
resulting splitting, $2|\Pi^{\times}_{sd}(q^*)|$. Writing the
off-diagonal coupling as $|\Pi^{\times}_{sd}|=c\,(v_Fq)^2/2\Delta$, the dimensionless prefactor $c$ follows from Eq.~\eqref{eq:final} evaluated at the crossing. Using the parameters listed for the Ba-122 superconductor above, a crossing located at
$v_Fq^*\sim\Delta$ would correspond to
$q^*\sim2\times10^{-2}$~\AA$^{-1}$ and a splitting of order
$c\times10$~meV, i.e.\ in the meV range for $c\sim0.1$. Two effects push
in the same direction as disorder increases: the amplitude branch
flattens, and, disorder being pair-breaking for the anisotropic
subdominant channel, it simultaneously weakens the effect from $\lambda_d$ raising
$\Omega_{BS}(0)$ toward the edge thus reducing the numerator of
Eq.~\eqref{qstar}. What decides observability is the competition between this narrowing of the gap to be closed and the broadening of the BS resonance by the same pair-breaking scattering \cite{Hirschfeld2011,Silaev2019}. Establishing whether the window in $\tau\Delta$ where $\alpha_{\mathrm{SH}}< \alpha_{\mathrm{BS}}$ overlaps with the window where the BS mode remains well defined requires impurity-dressed amplitude and phase sectors in both pairing channels -- a diffusive-limit calculation that is beyond the clean framework of this paper and is the subject of separate work.

We emphasize, finally, that the selection rule itself is unaffected by disorder. The Higgs--BS coupling remains forbidden by particle--hole symmetry and is switched on only at $O(\Delta/\veps_F)$, with the $(v_Fq)^2\cos2\phi_{\bq}$ angular structure and the odd-frequency form protected by symmetry rather than by the clean-limit kinematics and only its coefficient would change. The calculation presented here therefore supplies the coupling that would control the splitting in precisely the regime where a resonance can occur.

\section{Conclusions}

We have computed the finite-momentum pair-susceptibility matrix of a
two-dimensional spin-singlet superconductor with a leading $s$-wave and a subleading $d$-wave pairing channel, including the long-range Coulomb interaction, and used it to determine whether and how the $s$-wave amplitude (Higgs) mode and the $d$-wave Bardasis--Schrieffer (BS) mode interact.

Our first result is a selection rule. The two excitations reside in
different sectors of the order parameter --- the Higgs in the amplitude ($A_{1g}$) sector, the BS in the phase ($B_{1g}$) sector --- and within a particle--hole symmetric theory they decouple at \emph{all} momenta. Finite momentum alone does not lift the prohibition; it generates only intra-sector
mixing. The coupling is switched on exclusively by the leading
particle--hole--asymmetric correction, of order $\Delta/\veps_F$. We
obtained it in closed form,
$\Pi^{\times}_{sd}(\bq,\Omega)\propto(\Delta/\veps_F)(v_Fq)^2\Omega
\cos2\phi_{\bq}$ times a universal frequency profile, and confirmed each
feature --- the $q^2$ onset, the $\cos2\phi_{\bq}$ angular law, the
$\Delta/\veps_F$ scaling and the odd-frequency structure --- by an
independent evaluation of the corresponding loop diagram.

Our second result is that this coupling does not generate a resonance in a clean superconductor, but does so for the disordered one. The Higgs of a clean superconductor carries no sub-gap spectral weight: it is a
resonance attached to the pair-breaking threshold, which disperses as
$\Omega^2_{\rm edge}=4\Delta^2+(v_Fq)^2$. Any collective state bound below
that continuum disperses more slowly, and the separation between the two
branches therefore grows with momentum, Eq.~\eqref{nogo}, rather than
closing. There is no avoided crossing and no splitting. The conclusion is
kinematic: it follows from the dispersion of the threshold itself and is
independent of the magnitude of the coupling, of the material parameters,
and --- by the same argument --- applies to any sub-gap mode of a competing
pairing channel, including Leggett and mixed-symmetry BS modes.
What the coupling does produce is a transfer of amplitude character to the
BS mode, carried by the off-diagonal weight $A_{14}$, which is nonzero only
through the particle--hole asymmetry and therefore vanishes both at $\bq=0$
and along the nodal direction $\phi_{\bq}=\pi/4$. We find in addition
that the BS mode remains a sharp sub-gap pole at every momentum in the
clean limit: with an isotropic condensate the continuum is bounded below by
$2\Delta$ for all $\bq$, and the bound state, dispersing with
$\alpha_{BS}\simeq1/2$, approaches the continuum bottom with a binding
that falls below our numerical resolution for $v_Fq\gtrsim1.7\Delta$.

More broadly, these results identify particle-hole asymmetry as the
necessary ingredient coupling the amplitude and phase sectors -- a
selection rule that applies equally to the Higgs--plasmon and Higgs--Leggett
problems --- while showing that finite momentum cannot be used to bring the Higgs and a sub-gap collective mode into resonance in the clean limit. 

The natural extension is the regime of moderate disorder, $\tau\Delta\sim 1$. There the amplitude resonance detaches from the pair-breaking edge and its dispersion softens through zero on its way to the downward-dispersing dirty limit~\cite{Nosov2025}, so that the crossing condition $\alpha_{SH}<\alpha_{BS}$ of Eq.~\eqref{qstar} can be met and the coupling computed here would set the resulting splitting. Whether that window overlaps with the one in which the subdominant channel survives its own pair-breaking suppression is the question that decides if a genuine Higgs--BS resonance exists at all.

\begin{acknowledgments}
The work of SA and MD was financially supported by the National Science Foundation Grant No. DMR-2400484. Two of us (YB and MD) have performed the main part of this work at the Aspen Center for Physics, which is supported by the National Science Foundation Grant No. PHY-2210452. MD also thanks the Niels Bohr Institute, where part of this work was completed, for its hospitality. The authors acknowledge the use of Anthropic's Claude for cross-checking the analytical derivations, for independent numerical verification of the results, and for figure preparation and manuscript checking. All outputs of Claude were checked and validated by the authors.
\end{acknowledgments}

\section*{Data availability} The data that support the findings of this article were generated by numerical evaluation of the analytical expressions derived herein [Eqs.~\eqref{LargeHiggsBS} and \eqref{Aspec},
Appendices~\ref{AppendixA} and \ref{AppendixC}], with all model parameters specified in Sec. IV, and can be regenerated in full from the information given in the paper. For this reason, the authors did not deposit them in a public repository. The data and the scripts that produce Figs.~1 and 3--5 are available from the authors upon reasonable request. Figure 2 is a schematic and involves no data.

\begin{appendix}
\section{Corrections to quasiclassical propagator}\label{AppendixA}
We proceed with the calculation of the correction to the retarded and advanced components of $\check{g}$. These corrections are generated by the longitudinal component of the pairing field, which we represent as
\beg\label{deltaDelta}
\delta\hat{\Delta}_\bn(\br,t)=\left(i\hat{\tau}_2\right)\delta\Delta_\bn^L(\bq,\nu)e^{i(\bq\br-\nu t)}.
\en
Here we explicitly allow for the order parameter to fluctuate into either $s$-wave or $d$-wave channel. 
\subsection{Correction to the retarded and advanced components}
We write:
\beg\label{g2}
\delta\hat{g}(\bn\eps;\br t)=\delta\hat{g}(\bn\eps;\bk\omega)e^{2i(\bk\br-\omega t)},
\en
so that $\bq=2\bk$ and $\nu=2\omega$. In what follows, we will omit $R(A)$ superscripts for brevity. 
Equation for the function $\delta\hat{g}^{R(A)}(\bn\eps;\bk\omega)$ reads
\beg\label{Eq4g2a}
\begin{split}
&[\eps\hat{\tau}_3+\hat\Delta,\delta\hat{g}]+\omega\left\{\hat{\tau}_3,\delta\hat{g}\right\}-2{v}_F(\bn\bk)\delta\hat{g}\\&=-\left[\delta\hat{\Delta}_\bn^L\stackrel{\circ},\hat{g}\right]
\end{split}
\en
and $\delta\hat{\Delta}_\bn^L=i\hat{\tau}_2\delta\Delta_\bn^L$.
Given \eqref{deltaDelta} the commutator on the right hand side of this equation is given by
\beg\label{CommLHS}
\left[\delta\hat{\Delta}_\bn^L\stackrel{\circ},\hat{g}\right]=\delta\hat{\Delta}_\bn^L\hat{g}_{\eps-\omega}-\hat{g}_{\eps+\omega}\delta\hat{\Delta}_\bn^L
\en
and $\hat{g}_\eps$ denotes the quasiclassical propagator in equilibrium. 
For the expression entering on the left-hand side of \eqref{Eq4g2a} it obtains
\beg\label{re-writeg2}
\begin{split}
&[\eps\hat{\tau}_3+\hat\Delta_\bn,\delta\hat{g}]+{\omega}\left\{\hat{\tau}_3,\delta\hat{g}\right\}-2{v}_F(\bn\bk)\delta\hat{g}\\&=\eta_{\eps_+}\hat{g}_{\eps_+}\delta\hat{g}(\bn\eps;\bk\omega)-\eta_{\eps_{-}}\delta\hat{g}(\bn\eps;\bk\omega)\hat{g}_{\eps_{-}}
\\&-2{v}_F(\bn\bk)\delta\hat{g}(\bn\eps;\bk\omega)
\end{split}
\en
and $\eps_{\pm}=\eps\pm\omega$.
This expression can be further simplified using the normalization condition \eqref{norm}:
\beg\label{normg2RA}
\hat{g}_{\eps_+}\delta\hat{g}(\bn\eps;\bk\omega)+\delta\hat{g}(\bn\eps;\bk\omega)\hat{g}_{\eps_{-}}=0.
\en
Then the solution for the retarded and advanced components of $\delta\check{g}$ is
\beg\label{g2RAFinal}
{\delta\hat{g}^{R(A)}=\frac{\hat{\Lambda}_{\eps}^{R(A)}(\bk,\omega)\left(\delta\hat{\Delta}_\bn^L-\hat{g}_{\eps+\omega}^{R(A)}\delta\hat{\Delta}_\bn^L\hat{g}_{\eps-\omega}^{R(A)}\right)}{\left(\eta_{\eps+\omega}^{R(A)}+\eta_{\eps-\omega}^{R(A)}\right)^2-4{v}_F^2(\bn\bk)^2}}
\en
Here 
\beg\label{Gkw}
\hat{\Lambda}_{\eps}^{R(A)}=\left(\eta_{\eps+{\omega}}^{R(A)}+\eta_{\eps-{\omega}}^{R(A)}\right)\hat{\tau}_0+2{v}_F(\bn\bk)\hat{g}_{\eps+{\omega}}^{R(A)}.
\en
These expressions coincide with those reported earlier \cite{Kazi2026}. They can also be easily generalized for the case when external field is present. 
\subsection{Correction to the Keldysh component}
Equation for the $\delta\hat{g}^K$ is of course the same as \eqref{Eq4g2a}:
\beg\label{Eq4g2Ka}
\begin{split}
[\eps\hat{\tau}_3+\hat\Delta_\bn,\delta\hat{g}^K]+\omega\left\{\hat{\tau}_3,\delta\hat{g}^K\right\}&-2{v}_F(\bn\bk)\delta\hat{g}^K\\&=-\left[\delta\hat{\Delta}_\bn^L\stackrel{\circ},\hat{g}^K\right].
\end{split}
\en
The solution for $\hat{g}_2^K$ is different from $\hat{g}_2^{R(A)}$ because it satisfies the different normalization condition:
\beg\label{norm4g2K}
\begin{split}
&\hat{g}_{\eps_{+}}^R\delta\hat{g}^K+\hat{g}_{\eps_{+}}^K\delta\hat{g}^A+\delta\hat{g}^K\hat{g}_{\eps_{-}}^A+\delta\hat{g}^R\hat{g}_{\eps_{-}}^K=0.
\end{split}
\en
We look for the solution of Eq. \eqref{Eq4g2Ka} in the following form
\beg\label{g2Kansatz}
{\delta\hat{g}^K=\delta\hat{g}^Rt_{\eps-\omega}-t_{\eps+\omega}\delta\hat{g}^A+\delta \hat{g}_2^K}
\en
and $t_\eps=\tanh(\eps/2T)$.
Inserting this ansatz into \eqref{norm4g2K} yields
\beg\label{norm4dg2K}
\hat{g}_{\eps+\omega}^R\delta\hat{g}^K+\delta\hat{g}_2^K\hat{g}_{\eps-\omega}^A=0.
\en
Using \eqref{g2Kansatz} we derive the following equation for the function $\delta\hat{g}^K$
\beg\label{Eq4dg2K}
\begin{aligned}
&[\eps\hat{\tau}_3+\hat\Delta_\bn,\delta\hat{g}^K]+\omega\left\{\hat{\tau}_3,\delta\hat{g}^K\right\}-2{v}_F(\bn\bk)\delta\hat{g}_2^K\\&=\left[\delta\hat{\Delta}_\bn^L\hat{g}_{\eps-\omega}^A-\hat{g}_{\eps+\omega}^R\delta\hat{\Delta}_\bn^L\right](t_{\eps-\omega}-t_{\eps+\omega})
\end{aligned}
\en
Introducing function
\beg\label{LamKRK}
\begin{split}
\hat{\Lambda}_{\bn\eps}^K(\bk\omega)&=\left(\eta_{\bn\eps+\omega}^R+\eta_{\bn\eps-\omega}^A\right)\hat{\tau}_0+2v_F(\bn\bk)\hat{g}_{\bn\eps+\omega}^R, 
\end{split}
\en
the resulting expression for the function $\delta\hat{g}^K$ reads
\beg\label{dg2KFinal}
\delta\hat{g}^K=\frac{\hat{\Lambda}_{\eps}^{K}\left(\delta\hat{\Delta}_\bn^L-\hat{g}_{\eps_{+}}^R\delta\hat{\Delta}_\bn^L\hat{g}_{\eps_{-}}^A\right)
(t_{\eps_{+}}-t_{\eps_{-}})}{\left(\eta_{\eps_{+}}^{R}+\eta_{\eps_{-}}^{A}\right)^2-4{v}_F^2(\bn\bk)^2}.
\en
We can now use these expressions to derive an expression for the longitudinal pair susceptibility.

\subsection{Longitudinal (Schmid-Higgs) pairing susceptibility}
\paragraph{Single pairing channel}
In this section, we will derive the expression for the susceptibility of the amplitude Schmid-Higgs (SH) mode for the clean $s$ or $d$-wave superconductor. 
The expression for the SH susceptibility can be derived from the self-consistency equation 
\beg\label{SelfL}
\begin{aligned}
\delta\Delta^L(\bk,\omega)&=\frac{\lambda}{2}\int\limits_0^{2\pi}\frac{d\theta_\bn}{2\pi}{\gamma}(\theta_\bn)\int\limits_{-\infty}^\infty{d\eps}\textrm{Tr}\left\{-i\hat{\tau}_2\delta\hat{g}^K\right\},
\end{aligned}
\en
where $\lambda$ is the dimensionless pairing strength of the corresponding channel. The expression for 
$\delta\Delta_\bn^L(\bk\omega)$ can also be simplified by substituting
\beg\label{Lamkw}
\begin{aligned}
&\hat{\Lambda}_{\eps}^{R(A)}\left({\bk}{\omega}\right)\to\eta_{\eps+{\omega}}^{R(A)}+\eta_{\eps-{\omega}}^{R(A)}, \\
&\hat{\Lambda}_{\eps}^{K}\left({\bk}{\omega}\right)\to\eta_{\eps+{\omega}}^{R}+\eta_{\eps-{\omega}}^{A},
\end{aligned}
\en
since the odd-in-powers of $\bn$ term in functions  $\hat{\Lambda}_{\eps}^{a}\left({\bk}{\omega}\right)$ will drop out upon integration over $\theta_\bn$.
Inserting the expressions above into the self-consistency equation \eqref{Self} yields the linear consistency relation in the form $\chi_{\textrm{SH}}^{-1}(\bq,\Omega)\delta\Delta^L=0$ where $\chi_{\textrm{SH}}^{-1}(\bq,\Omega)$ has a physical meaning of the inverse
longitudinal susceptibility, and it is given by
\begin{widetext}
\beg\label{chiSHdwave}
\begin{aligned}
\chi_{\textrm{SH}}^{-1}(\bq,\Omega)=-\frac{1}{\lambda}&+\int\limits_{-\omega_D}^{\omega_D}d\eps\int\limits_{0}^{2\pi}\frac{d\theta_\bn}{2\pi}\left\{
\frac{\left(\eta_{\eps+\Omega/2}^{R}+\eta_{\eps-\Omega/2}^{A}\right){\cal A}^K(\eps_+,\eps_-)(t_{\eps+\Omega/2}-t_{\eps-\Omega/2})}{\left(\eta_{\eps+\Omega/2}^{R}+\eta_{\eps-\Omega/2}^{A}\right)^2-{v}_F^2(\bn\bq)^2}\right.\\&\left.+\frac{\left(\eta_{\eps+\Omega/2}^{R}+\eta_{\eps-\Omega/2}^{R}\right){\cal A}^R(\eps_+,\eps_-)t_{\eps-\Omega/2}}{\left(\eta_{\eps+\Omega/2}^{R}+\eta_{\eps-\Omega/2}^{R}\right)^2-{v}_F^2(\bn\bq)^2}-
\frac{\left(\eta_{\eps+\Omega/2}^{A}+\eta_{\eps-\Omega/2}^{A}\right){\cal A}^A(\eps_+,\eps_-)t_{\eps+\Omega/2}}{\left(\eta_{\eps+\Omega/2}^{A}+\eta_{\eps-\Omega/2}^{A}\right)^2-{v}_F^2(\bn\bq)^2}\right\}.
\end{aligned}
\en
\end{widetext}
Here ${\cal A}^K(\eps,\eps')=1+g_{\eps}^Rg_{\eps'}^A+f_{\eps}^Rf_{\eps'}^A$, ${\cal A}^b(\eps,\eps')=1+g_{\eps}^bg_{\eps'}^b+f_{\eps}^bf_{\eps'}^b$ ($b=R,A$), $\eps_{\pm}=\eps\pm\Omega/2$, $\Omega=2\omega$, the dimensionless coupling constant is given by
\beg\label{Coupling}
\frac{1}{\lambda}=\frac{1}{\Delta}\int\limits_{-\omega_D}^{\omega_D}d\eps\int\limits_0^{2\pi}\frac{d\theta_\bn}{2\pi}\left(f_{\eps}^R-f_{\eps}^A\right)t_\eps,
\en
where $\Delta$ is the value of the order parameter in equilibrium. 
It is worth noting that in the limit $\bq=0$ and taking ${\cal Y}(\theta_\bn)=1$ we recover previously derived  expression for the Schimd-Higgs susceptibility for the $s$-wave superconductor. Note also that while the integrals in Eqs. \eqref{chiSHdwave} and \eqref{Coupling} need to be
cut off at a Debye frequency $\omega_D$, being taken together they yield
a UV convergent integral. Thus expression for the inverse
susceptibility, $\chi_{\textrm{SH}}^{-1}(\bq,\Omega)$, is, in fact, cutoff independent. 
\paragraph{Two pairing channels.}
Generally, the pairing interaction can be written in separable form over an orthonormal basis of
Fermi-surface harmonics $\{{\cal Y}_a(\theta_\bn)\}$, $\langle{{\cal Y}_a{\cal Y}_b}\rangle_{\bn}=\delta_{ab}$,
\begin{equation}
V(\bn,\bn')=-\sum_a \lambda_a\,{\gamma}_a(\theta_\bn)\,{\gamma}_a(\theta_{\bn'}).
\end{equation}
For a tetragonal system, for example, the relevant labels are
$A_{1g}$ (isotropic-$s$/extended-$s$, ${\gamma}_s=1$),
$B_{1g}$ ($d_{x^2-y^2}$, ${\gamma}_d=\sqrt2\cos2\phi$),
$B_{2g}$ ($d_{xy}$), etc. Each channel keeps its own bare coupling $\lambda_a$, and the gap equation
fixes only the dominant one. As a result, one finds \eqref{SH2x2} in the main text.
Note that in this case the matrix elements of $[\chi_{\textrm{SH}}^{-1}]_{ab}$ ($a,b=s,d$) are computed using the same formula as above with the angular integral weighted by the corresponding pairing form factors. 
\subsection{Transverse pair susceptibility}
The expression for the transverse susceptibility is obtained in full analogy with the calculation which led to \eqref{chiSHdwave}, with $\delta\hat{g}^K$ computed by solving \eqref{Eq4g2RAT}.
$\chi_{\textrm{AB}}^{-1}(\bq,\Omega)$ is defined as
\begin{widetext}
\beg\label{chiABdwave}
\begin{split}
&\chi_{\textrm{AB}}^{-1}(\bq,\Omega)=-\frac{1}{\lambda}+\int\limits_{0}^{2\pi}{\gamma}^2(\theta_\bn)\frac{d\theta_\bn}{2\pi}\int\limits_{-\omega_D}^{\omega_D}d\eps\left\{
\frac{\left(\eta_{\eps+\Omega/2}^{R}+\eta_{\eps-\Omega/2}^{A}\right){A}^K(\eps_+,\eps_-)(t_{\eps+\Omega/2}-t_{\eps-\Omega/2})}{\left(\eta_{\eps+\Omega/2}^{R}+\eta_{\eps-\Omega/2}^{A}\right)^2-{v}_F^2(\bn\bq)^2}\right.\\&\left.+\frac{\left(\eta_{\eps+\Omega/2}^{R}+\eta_{\eps-\Omega/2}^{R}\right){A}^R(\eps_+,\eps_-)t_{\eps-\Omega/2}}{\left(\eta_{\eps+\Omega/2}^{R}+\eta_{\eps-\Omega/2}^{R}\right)^2-{v}_F^2(\bn\bq)^2}-
\frac{\left(\eta_{\eps+\Omega/2}^{A}+\eta_{\eps-\Omega/2}^{A}\right){A}^A(\eps_+,\eps_-)t_{\eps+\Omega/2}}{\left(\eta_{\eps+\Omega/2}^{A}+\eta_{\eps-\Omega/2}^{A}\right)^2-{v}_F^2(\bn\bq)^2}\right\},
\end{split}
\en
\end{widetext}
where $\eps_{\pm}=\eps\pm\Omega/2$ and 
\beg\label{AKRAT}
\begin{aligned}
{A}_\bn^{R(A)}(\eps,\eps')=g_{\bn\eps}^{R(A)}g_{\bn\eps'}^{R(A)}-f_{\bn\eps}^{R(A)}f_{\bn\eps'}^{R(A)}+1, \\
{A}_\bn^{K}(\eps_{+},\eps_{-})=g_{\bn\eps}^{R}g_{\bn\eps'}^{A}-f_{\bn\eps}^{R}f_{\bn\eps'}^{A}+1.
\end{aligned}
\en
The definition of the function $\rho_{\textrm{a}}({\bq,\Omega})$ which accounts for the hybridization of the leading (${\mathrm a}=s$) or sub-leading (${\mathbf a}=d$) phase modes with the Coulomb fluctuations is 
\begin{widetext}
\beg\label{zetaetc}
\begin{aligned}
&\rho_{\textrm{a}}({\bq,\omega})=\int\limits_{0}^{2\pi}\frac{d\theta_\bn}{2\pi}{\gamma}_{\textrm{a}}(\theta_\bn)\int\limits_{-\omega_D}^{\omega_D}d\eps\left\{
\frac{\left(\eta_{\eps+\Omega/2}^{R}+\eta_{\eps-\Omega/2}^{A}\right){B}^K(\eps_+,\eps_-)(t_{\eps+\Omega/2}-t_{\eps-\Omega/2})}{\left(\eta_{\eps+\Omega/2}^{R}+\eta_{\eps-\Omega/2}^{A}\right)^2-{v}_F^2(\bn.\bq)^2}\right.\\&\left.+\frac{\left(\eta_{\eps+\Omega/2}^{R}+\eta_{\eps-\Omega/2}^{R}\right){B}_\bn^R(\eps_+,\eps_-)t_{\eps-\Omega/2}}{\left(\eta_{\eps+\Omega/2}^{R}+\eta_{\eps-\Omega/2}^{R}\right)^2-{v}_F^2(\bn.\bq)^2}-
\frac{\left(\eta_{\eps+\Omega/2}^{A}+\eta_{\eps-\Omega/2}^{A}\right){B}^A(\eps_+,\eps_-)t_{\eps+\Omega/2}}{\left(\eta_{\eps+\Omega/2}^{A}+\eta_{\eps-\Omega/2}^{A}\right)^2-{v}_F^2(\bn.\bq)^2}\right\}.
\end{aligned}
\en
Last but not least, we provide the definition of the Coulomb response function $\chi_{\textrm{CG}}^{-1}(\bq,\Omega)$:
\beg\label{chiCC}
\begin{aligned}
&\chi_{\textrm{CG}}^{-1}(\bq,\omega)=\int\limits_{0}^{2\pi}\frac{d\theta_\bn}{2\pi}\int\limits_{-\infty}^{\infty}d\eps\left\{
\frac{\left(\eta_{\eps+\Omega/2}^{R}+\eta_{\eps-\Omega/2}^{A}\right)\widetilde{A}^K(\eps_+,\eps_-)(t_{\eps+\Omega/2}-t_{\eps-\Omega/2})}{\left(\eta_{\eps+\Omega/2}^{R}+\eta_{\eps-\Omega/2}^{A}\right)^2-{v}_F^2(\bn.\bq)^2}\right.\\&\left.+\frac{\left(\eta_{\eps+\Omega/2}^{R}+\eta_{\eps-\Omega/2}^{R}\right)\widetilde{A}^R(\eps_+,\eps_-)t_{\eps-\Omega/2}}{\left(\eta_{\eps+\Omega/2}^{R}+\eta_{\eps-\Omega/2}^{R}\right)^2-{v}_F^2(\bn.\bq)^2}-
\frac{\left(\eta_{\eps+\Omega/2}^{A}+\eta_{\eps-\Omega/2}^{A}\right)\widetilde{A}^A(\eps_+,\eps_-)t_{\eps+\Omega/2}}{\left(\eta_{\eps+\Omega/2}^{A}+\eta_{\eps-\Omega/2}^{A}\right)^2-{v}_F^2(\bn.\bq)^2}\right\}.
\end{aligned}
\en
\end{widetext}
Here $\eps_\pm=\eps\pm\frac{\Omega}{2}$.
In the expressions above, we are using the following functions
\beg\label{tilAKRA}
\begin{aligned}
&\widetilde{A}^{R(A)}(\eps,\eps')=g_{\eps}^{R(A)}g_{\eps'}^{R(A)}-f_{\eps}^{R(A)}f_{\eps'}^{R(A)}-1, \\
&\widetilde{A}^{K}(\eps,\eps')=g_{\eps}^{R}g_{\eps'}^{A}-f_{\eps}^{R}f_{\eps'}^{A}-1, \\
\end{aligned}
\en
and 
\beg\label{tilAKRA2}
\begin{aligned}
&{B}^K(\eps,\eps')=g_\eps^Rf_{\eps'}^A-f_\eps^Rg_{\eps'}^A, \\ 
&{B}^{R(A)}(\eps,\eps')=g_\eps^{R(A)}f_{\eps'}^{R(A)}-f_\eps^{R(A)}g_{\eps'}^{R(A)}.
\end{aligned}
\en

\section{Calculation of the hybridization function $\Pi_{\mathrm{ab}}^{\times}(\bq,i\Omega_l)$}\label{AppendixC}
In this Section we provide the details on the calculation of the response function describing hybridization between the amplitude and phase modes.
\paragraph{Expansion in powers of small momentum}
We start with Eq. \eqref{SummedUp} in the main text. Expanding the expression under the integral yields
\beg\label{Expand1}
\begin{aligned}
\frac{\frac{\xi_{+}}{E_{+}}+\frac{\xi_{-}}{E_{-}}}{(i\Omega_l)^2-(E_{+}+E_{-})^2}&\approx\frac{2\xi_\bk}{E_\bk[(i\Omega_l)^2-4E_\bk^2]}
\\&-u^2\,\mathcal K_2(\xi)+O(u^4),
\end{aligned}
\en
Here $u=(v_F/2)\sqrt{1+\xi/\veps_F}(\bn\bq)$ and we introduced function
\beg\label{K2}
\mathcal K_2(\xi)=\frac{3\Delta^2\xi_\bk}{E_\bk^5[(i\Omega_m)^2-4E_\bk^2]}-\frac{8\Delta^2\xi_\bk}{E_\bk^3[(i\Omega_m)^2-4E_\bk^2]^2}
\en
and $E_\bk=(\xi_\bk^2+\Delta^2)^{1/2}$.
Now it remains to integrate over $\xi$. By making a standard substitution 
\beg\label{IntSub}
\int\frac{d^2\bk}{(2\pi)^2}(...)=\nu_F\int\limits_{-\veps_F}^\infty d\xi_\bk\int\frac{d\theta_\bk}{2\pi}(...).
\en
Given \eqref{Expand1} we see that the integrals over $\xi_\bk$ and $\theta_\bk$ disentangle. The angular integral for the diagonal components of $\Pi_{\mathrm{ab}}^{(\times)}$ gives $1/2$, while the for the off-diagonal components we find $\cos(2\theta_\bq)/2\sqrt{2}$. The lower limit of the integral over $\xi_\bk$ in the first term in \eqref{Expand1} can be extended to $-\infty$ and yields zero. The remaining energy integral projects out the particle-hole asymmetry: $\mathcal K_2$ is \emph{odd} in $\xi$, so with a constant density of states the
symmetric part $\int d\xi\,\mathcal K_2=0$, while the $\xi/\veps_F$ carried by
$u^2$ survives. It obtains
\begin{equation}
\begin{aligned}
\frac{\nu_F}{\veps_F}\!\int\limits_{-\veps_F}^\infty\! \xi d\xi\,\mathcal K_2(\xi)
=&\frac{\nu_F\Delta^2}{\veps_F}\int_{-\infty}^{\infty}\! \frac{\xi^2d\xi}{E^3[(i\Omega_l)^2-4E^2]}
\\&\times\left[\frac{3}{E^2}-\frac{8}{(i\Omega_l)^2-4E^2}\right].
\end{aligned}
\label{eq:calI}
\end{equation}
This is the microscopic origin of the $\Delta/\veps_F$ pre-factor.

The remaining integral can be computed as follows. Since we need to work with real frequencies, we replace $i\Omega_l=\Omega+i0$ and introduce the dimensionless parameter $z=\Omega/2\Delta$ and a new integration variable via $\xi=\Delta\sinh t$ followed by $w=\tanh t$. Then we find
\begin{equation}\label{GzInt}
\frac{\nu_F}{\veps_F}\!\int\limits_{-\veps_F}^\infty\! \xi d\xi\,\mathcal K_2(\xi)=\frac{2\nu_F}{\veps_F\Delta^2}G(z),
\en
where 
\beg\label{DefineGz}
\begin{aligned}
G(z)&=\int\limits_0^1\!\,\frac{w^2(1-w^2)\,(3z^2-5-3z^2w^2)dw}{4\,(z^2-1-z^2w^2)^2}\\
&=\frac{1}{4z^2}-\frac{\arcsin z}{4z^3\sqrt{1-z^2}}.
\end{aligned}
\en
Multiplying this function by the factor from the angular integrals defines function $G_{\textrm{ab}}$ in the main text. 

\paragraph{Elimination of the Coulomb field: the screened $4\times4$ problem}

The full fluctuation problem couples the four order-parameter components
$\delta\vec\Delta=(\delta\Delta_s^L,\,\delta\Delta_d^L,\,
\delta\Delta_s^T,\,\delta\Delta_d^T)$ to the scalar potential
$\delta\Phi$. Collecting the four gap self-consistency equations and the
Poisson equation yields the $5\times5$ system
\beg\label{FiveByFive}
\begin{aligned}
&\sum_{\textrm{b}=1}^{4}\hat{M}_{\textrm{ab}}\,\delta\Delta_{\textrm{b}}
      + v_{\textrm{a}}\,\delta\Phi=0, \\
&\sum_{\textrm{a}=1}^{4} v_{\textrm{a}}^{*}\,\delta\Delta_{\textrm{a}}
      + D_\Phi(\bk,\omega)\,\delta\Phi=0,
\end{aligned}
\en
that is, $\check{\mathcal X}^{-1}=\left(\begin{smallmatrix}
\hat M & \vec v\\[2pt] \vec v^{\dagger} & D_\Phi\end{smallmatrix}\right)$,
with the potential's inverse propagator
\beg\label{DPhi}
D_\Phi(\bk,\omega)=4+\chi^{-1}_{\textrm{CG}}(\bk,\omega)
+\frac{4k}{k_\mathrm{TF}}.
\en
Here we introduced the Thomas-Fermi momentum $k_\mathrm{TF}={2\pi\nu_Fe^2}$ and the coupling column
$\vec{v}=(\kappa\Pi_{s\varphi},\;\kappa\Pi_{d\varphi},\;
i\rho_s,\;i\rho_d)$,
where $\kappa=\Delta/\veps_F$, the $\rho_{s,d}$ are the particle-hole--even
pair--density couplings [$\rho_s=O(1)$,
$\rho_d\propto(v_Fq)^2\cos2\phi_\bq$] and $\kappa\Pi_{a\varphi}$ the
particle-hole--odd amplitude--density bubbles. The Poisson row is the
Hermitian conjugate $\vec v^{\dagger}$ of the gap-equation column---each
coupling is odd in frequency, so the phase entries flip sign,
$i\rho_a\to-i\rho_a$. The charge therefore couples \emph{antisymmetrically}
to the order parameter, the same non-symmetry carried by the
amplitude--phase block $\Pi^{\times}$, in agreement with
Eqs.~\eqref{LargeTSystem} and \eqref{LargeHiggsBS}. Solving the second line
of \eqref{FiveByFive} for $\delta\Phi$ and substituting into the first
gives the exact Schur complement
\beg\label{Schur4}
\widetilde{M}_{\textrm{ab}}=M_{\textrm{ab}}
-\frac{v_{\textrm{a}}\,v_{\textrm{b}}^{*}}{D_\Phi(\bk,\omega)}.
\en

Two properties of \eqref{Schur4} organize the analysis. First, in the
regime of interest, $k\sim q^*\ll k_{TF}$, the explicit Coulomb kernel in
\eqref{DPhi} is negligible and
\beg\label{DPhiNeutral}
D_\Phi(\bk,\omega)\simeq 4+\chi^{-1}_{\textrm{CG}}(\bk,\omega),
\en
so the electron charge drops out of the sub-gap problem entirely: $e^2$ is
needed only to place the plasmon itself. Second, the antisymmetric column
$\vec v^{\dagger}$ fixes the sign of every correction, and these organize
by sector:
\begin{widetext}
\begin{center}
\begin{tabular}{lll}
\hline
block & correction & order \\
\hline
phase--phase ($a,b\in\{3,4\}$) & $-\rho_a\rho_b/D_\Phi$ & $O(1)$ \\
amplitude--phase ($a\in\{1,2\}$, $b\in\{3,4\}$) &
$+i\kappa\,\Pi_{a\varphi}\rho_b/D_\Phi$ & $O(\kappa)$ \\
amplitude--amplitude ($a,b\in\{1,2\}$) &
$-\kappa^2\Pi_{a\varphi}\Pi_{b\varphi}/D_\Phi$ & $O(\kappa^2)$, dropped \\
\hline
\end{tabular}
\end{center}
\end{widetext}
The $O(1)$ phase-sector correction implements the screening physics
established in the single-channel analysis: $-\rho_s^2/D_\Phi$ lifts the
Anderson--Bogoliubov zero out of the sub-gap window (plasmonization),
while the $\rho_d$-containing entries dress the Bardasis--Schrieffer
diagonal and the transverse mixing at $O\big((v_Fq)^4\big)$ and
$O\big((v_Fq)^2\big)$ respectively---numerically at the sub-percent level
at the momenta of interest. The $O(\kappa)$ correction to the Higgs--BS
element,
\beg\label{induced}
\Pi^{\times,\textrm{ind}}_{sd}
=+\,\frac{i\kappa\,\Pi_{s\varphi}\,\rho_d}{4+\chi^{-1}_{\textrm{CG}}},
\en
is the potential-mediated contribution and enters at the same order,
$\kappa(\Omega/\Delta)(v_Fq)^2\cos2\phi_\bq$, as the direct cross-bubble
$\Pi^{\times}_{sd}$ (the two carry correlated signs, and only their sum is
physical). The screened matrix \eqref{Schur4} with the elements computed
in the preceding sections, supplemented by the five particle-hole--odd
bubbles $(\Pi^{\times}_{ss},\Pi^{\times}_{dd},\Pi^{\times}_{sd},
\Pi_{s\varphi},\Pi_{d\varphi})$ and by $\rho_s$, $\rho_d$, therefore
constitutes the complete input of the Higgs--BS hybridization problem. Its
determinant resums the direct and all indirect coupling paths consistently
to $O(\kappa)$.

\end{appendix}

\bibliography{bslightq}

\end{document}